\documentstyle[preprint,pre,aps,graphicx]{revtex}
\tightenlines

\newcommand{\eij}{{\bf\hat{e}}_{ij}}

\begin{document}
\draft

\title{Spinodal decomposition of off-critical quenches with a viscous
  phase using dissipative particle dynamics in two and three spatial
  dimensions}

\author{Keir E. Novik\thanks{Present address: Centre for Computational
  Science, Department of Chemistry, Queen Mary \& Westfield College,
  University of London, Mile End Road, London, E1 4NS, United Kingdom.
  Electronic address: K.E.Novik@qmw.ac.uk}}
\address{Cavendish Laboratory, University of Cambridge,
  Madingley Road, Cambridge, CB3 0HE, United Kingdom}

\author{Peter V. Coveney\thanks{Electronic address:
  P.V.Coveney@qmw.ac.uk}}
\address{Centre for Computational Science, Department of Chemistry,
  Queen Mary \& Westfield College, University of London, Mile End
  Road, London, E1 4NS, United Kingdom}

\date{Accepted by {\em Phys.\ Rev.\ E\/} on October 11, 1999}

\maketitle

\begin{abstract}
  We investigate the domain growth and phase separation of
  hydrodynamically-correct binary immiscible fluids of differing
  viscosity as a function of minority phase concentration in both two
  and three spatial dimensions using dissipative particle dynamics.
  We also examine the behavior of equal-viscosity fluids and compare
  our results to similar lattice-gas simulations in two dimensions.
\end{abstract}

\section{Introduction}

Over the last few years, the dissipative particle dynamics (DPD) model
of complex fluids has received considerable attention.  It has matured
from its somewhat arbitrary initial formulation into a model with a
solid theoretical basis.  Furthermore, it has been applied with
considerable success to a large number of computer simulations of
complex fluid systems such as colloidal suspensions, polymeric fluids,
spinodal decomposition of binary immiscible fluids, and amphiphilic
fluids.  DPD also looks promising for simulating multiphase flows and
flow in porous media, and is now considered a useful technique
alongside the other complex fluid algorithms: molecular dynamics,
lattice-gas automata, and techniques based on the lattice-Boltzmann
equation.  No single technique can yet be applied to all situations,
and each has different strengths and weaknesses.  While molecular
dynamics is in principle the most accurate microscopic approach, in
practice it is too slow in both its quantum (Car--Parinello) and
classical forms because of its excessive detail.  Discrete mesoscopic
methods developed from lattice-gas automata have had some success, but
they too have problems, such as lacking Galilean invariance.  The
traditional approach of continuum fluid dynamics has met with limited
success in modeling behavior on the length and time scales
characteristic of complex fluids.

In this paper we investigate phase separation from both symmetric
(critical) and off-critical quenches in binary immiscible fluids of
generally differing viscosity using the hydrodynamically-correct DPD
model.  Our motivation is to probe and extend knowledge of the
behavior of differing-viscosity fluids, of application, for example,
to the action of detergents and the extraction of oil from reservoir
rocks.  Phase separation in {\em equal-viscosity\/} binary immiscible
fluid systems has been simulated using a variety of techniques,
including DPD~\cite{cov-nov:binary,%
cov-nov:erratum,nov-cov:binary,novik:finite-difference,jury:scaling};
molecular dynamics~\cite{velasco:spinodal,velasco:off-critical,%
velasco:separation,laradji:spinodal}; Monte
Carlo~\cite{velasco:spinodal,kumar:novel}; cell dynamical systems
without hydrodynamics~\cite{bib:so} and with Oseen tensor
hydrodynamics~\cite{bib:so2}; time-dependent Ginzburg--Landau models
with~\cite{bib:fv,bib:vf,bib:walc,bib:lwac,corberi:early} and without
hydrodynamics~\cite{corberi:early,bib:ctg,chakrabarti:off-critical};
lattice-gas automata~\cite{bib:rk,rothman:spinodal,rothman:review,%
bib:bal,bib:appert,emerton:microemulsions,emerton:erratum,%
weig:minority}; and lattice-Boltzmann techniques~\cite{bib:acg,%
chen:late,osborn:lb,orlandini:structured,gonnella:lamellar,%
gonnella:droplet,wagner:breakdown}.  Spinodal decomposition of {\em
differing-viscosity\/} immiscible fluids has previously been simulated
in two dimensions by a time-dependent Ginzburg--Landau model without
hydrodynamics~\cite{sappelt:glassy}.  We discuss the effect of the
proportion of each phase on the scaling behavior in both two and three
spatial dimensions.  We also examine the behavior of equal-viscosity
fluids, comparing our results to similar lattice-gas simulations in
two dimensions~\cite{weig:minority}.

After describing our fluid model in the following section, we discuss
the expected temporal development of the characteristic size of the
separating domains in Section~\ref{temporal}.  We then describe our
method for calculating the characteristic domain size and its rate of
growth in Section~\ref{estimating}.  This is followed in
Sections~\ref{results} and~\ref{discussion} by information on the
simulations performed and a discussion of the results, and by some
conclusions in Section~\ref{conclusions}.  Finally, as an Appendix to
this paper, we make a few comments on the high-performance computing
aspects of this work.

\section{The Fluid Model}

In 1992, Hoogerbrugge and Koelman proposed dissipative particle
dynamics~\cite{hoogerbrugge:dpd} as a novel particulate model for the
simulation of complex fluid behavior.  DPD was developed in an attempt
to capture the best aspects of molecular dynamics and lattice-gas
automata.  It avoids the lattice-based problems of lattice-gas
automata, yet maintains an elegant simplicity and larger scale that
keeps the model much faster than molecular dynamics.  This simplicity
also makes DPD highly extensible, such as for including the
interactions of complex molecules or modeling flow in an arbitrary
number of spatial dimensions.  The key features of the model are that
the fluid is grouped into packets, termed ``particles'', and that mass
and momentum are conserved but energy is not.  Particles are normally
interpreted as representing a coarse-graining of the fluid, so that
each particle contains many molecules~\cite{hoogerbrugge:dpd,%
espanol:dpd,espanol:chain,espanol:coarse}.  Since the intrinsic time
scale of DPD represents the correlated motion of mesoscopic packets of
atoms or molecules, it is typically orders of magnitude larger than
the time scale of molecular dynamics~\cite{groot:bridging}.  Particle
positions and momenta are real variables, and are not restricted to a
grid.

Espa{\~n}ol and Warren's analysis~\cite{espanol:dpd} showed that the
original DPD model does not satisfy detailed balance, so the
equilibrium states (if they exist) cannot be simply characterized.
Detailed balance is the condition equating the rates of forward and
backward transition probabilities in a dynamical system, and is a
sufficient (but not necessary) condition guaranteeing that the system
has a (Gibbsian) equilibrium
state~\cite{gardiner:stochastic,risken:fokker}.  Espa{\~n}ol and
Warren formulated a Fokker--Planck equation and equivalent set of
stochastic differential equations which lead to an analogous
continuous-time model,
\begin{equation}\label{SDE}
  \cases{%
    d{\bf p}_i = \displaystyle\sum_{j \ne i} {\bf F}_{ij} dt =
    \displaystyle\sum_{j \ne i} \left[ {\bf F}^C_{ij} dt + 
      {\bf F}^D_{ij} dt + {\bf F}^R_{ij} dW_{ij} \right]
    \smallskip\cr
    d{\bf x}_i = {\displaystyle {\bf p}_i \over\displaystyle m_i}
    dt. 
    } 
\end{equation}
In these equations, ${\bf p}_i$, ${\bf x}_i$, and $m_i$ denote the
momentum, position, and mass of particle $i$.  ${\bf F}^C_{ij}$ is a
conservative force acting between particles $i$ and $j$, while ${\bf
  F}^D_{ij}$ and ${\bf F}^R_{ij}$ are the dissipative and random
forces.  $dW_{ij} = dW_{ji}$ are independent increments of a Wiener
process.  By It\^o calculus
\begin{equation}
  dW_{ij} dW_{kl} = \left( \delta_{ik}\delta_{jl} +
    \delta_{il}\delta_{jk} \right) dt ,
\end{equation}
so $dW_{ij}$ is an infinitesimal of order $\frac{1}{2}$ and $dW_{ij}$
can be written $\theta_{ij}\sqrt{dt}$, where $\theta_{ij}=\theta_{ji}$
is a random variable with zero mean and unit
variance~\cite{gardiner:stochastic}.  Detailed balance is satisfied by
this continuous-time version of DPD with an appropriate choice for the
form of the forces~\cite{espanol:hydro}, and so equilibrium states are
guaranteed to exist and be Gibbsian.  To ensure that the associated
fluctuation--dissipation theorem holds, Espa{\~n}ol and Warren
suggested the forces assume the following forms~\cite{espanol:dpd}:
\begin{equation}\label{Fc}
  {\bf F}^C_{ij} = \alpha \omega_{ij} \eij,
\end{equation}
\begin{equation}\label{Fd}
  {\bf F}^D_{ij} = - \gamma \omega^2_{ij} \left( \eij \cdot{}
    {\bf v}_{ij} \right) \eij ,  
\end{equation}
and
\begin{equation}\label{Fr}
  {\bf F}^R_{ij} = \sigma \omega_{ij} \eij,
\end{equation}
where ${\bf v}_{ij} = ({\bf p}_i / m_i) - ({\bf p}_j / m_j)$ is the
difference in velocities of particles $j$ and $i$, $\eij$ is the unit
vector pointing from particle $j$ to particle $i$, and $\omega_{ij}$
is a weighting function depending only on the distance separating
particles $i$ and $j$.  The constants $\alpha$, $\gamma$, and $\sigma$
are chosen to reflect the relative importance of the conservative,
dissipative (viscous), and random components in the fluid of interest.
As a consequence of detailed balance and the fluctuation--dissipation
theorem, $\gamma$ and $\sigma$ are related to Boltzmann's constant
$k_B$ and the equilibrium temperature $T$ by
\begin{equation}\label{temperature}
  {\sigma^2 \over \gamma} = 2 k_B T.
\end{equation}
In order to remain as close as possible to the original DPD model,
Espa{\~n}ol and Warren chose the friction weight function to be
\begin{equation}\label{omega}
  \omega_{ij} = 1 - {\displaystyle r_{ij} \over\displaystyle r_c}
\end{equation}
within the constant cutoff length $r_c>0$ and zero otherwise, where
$r_{ij}$ is the distance between particles $i$ and $j$.  Adding
Eqs.~(\ref{Fc})--(\ref{Fr}) together, the total force is
\begin{equation}\label{dpd_force}
  {\bf F}_{ij} = \left[ \alpha - \gamma \omega_{ij} \left( \eij
      \cdot{}{\bf v}_{ij} \right) + {\sigma \theta_{ij} \over
      \sqrt{dt}} \right] \omega_{ij} \eij .
\end{equation}

Immiscible fluid mixtures exist because individual molecules attract
similar and repel dissimilar molecules.  The most common example of
such behavior arises in mixtures of oil and water below a critical
temperature.  The nonpolar, hydrophobic molecules of oil attract one
another through short-range van der Waals forces, while the polar
water molecules have complex, long-range hydrophilic attractions which
are dominated by electrostatic interactions, including hydrogen bonds.
At the atomistic level employed in molecular dynamics, such
interactions demand a detailed treatment.  However, to obtain
mesoscopic and macroscopic level descriptions using DPD, the
microscopic model can be drastically simplified.

In order to model immiscible fluids, the simplest modification to the
one-component dissipative particle dynamics algorithm is to introduce
a new variable, called the ``color'' by analogy with
Rothman--Keller~\cite{bib:rk,hoogerbrugge:dpd}.  For example, we could
choose red to represent oil and blue to represent water.  When two
particles of different color interact we increase the conservative
force, thereby increasing the repulsion.  That is,
\begin{equation}\label{ImmiscibleAlpha}
  \alpha \mapsto \alpha_{ij} = \cases{%
    \alpha_0 & if particles $i$ and $j$ are the same color \cr
    \alpha_1 & if particles $i$ and $j$ are different colors \cr
    }
\end{equation}
where $\alpha_0$ and $\alpha_1$ are constants with $0 \le \alpha_0 <
\alpha_1$.  As a consequence of mass and momentum conservation, the
Navier--Stokes equation is obeyed within a single-phase DPD fluid and
within regions of homogeneity of each of the two binary immiscible
fluid phases~\cite{hoogerbrugge:dpd,espanol:dpd,espanol:hydro,%
coveney:multicomponent}.  Likewise, detailed balance is also
preserved, at least in the limit of continuous
time~\cite{espanol:dpd,espanol:hydro,coveney:multicomponent}.

The immiscible fluids described above are identical to each other.
However, it is often the case that the two fluids in a mixture will
differ physically.  For example, oil and water typically have
different viscosities.  To model binary fluids with differing
viscosity we again adopt the simplest approach: labeling one of the
two phases (colors) as more viscous.  When two particles of the same
viscous color interact, the dissipative (viscous) force is increased;
in order to keep the temperature constant, we must correspondingly
decrease the random force according to Eq.~(\ref{temperature}), i.e.,
\begin{equation}
  \gamma \mapsto \gamma_{ij} = \cases{%
    \gamma_0 & if either particle $i$ or $j$ is not the viscous color \cr
    \gamma_1 & if both particles $i$ and $j$ are the viscous color. \cr
    }
\end{equation}
\begin{equation}
  \sigma \mapsto \sigma_{ij} = \cases{%
    \sigma_0 & if either particle $i$ or $j$ is not the viscous color \cr
    \sigma_1 & if both particles $i$ and $j$ are the viscous color, \cr
    }
\end{equation}
where 
\begin{equation}\label{temperature2}
  {\sigma_0^2 \over \gamma_0} = {\sigma_1^2 \over \gamma_1} = 2 k_B T.
\end{equation}

Our previous study of finite difference methods applicable to
simulations with non-conservative
forces~\cite{novik:finite-difference} indicated that the
finite-difference algorithm suggested by Groot and
Warren~\cite{groot:bridging} is a good choice for DPD.  Their method
is a variation on the familiar velocity-Verlet algorithm, adding a
momentum estimate before the force evaluation:
\begin{eqnarray}\label{groot-warren}
  {\bf x}_i(t + \Delta t) &=& {\bf x}_i(t) + \frac{\Delta t}{m_i}
  \left[ {\bf p}_i(t) + \frac{\Delta t}{2} {\bf f}_i(t)
  \right] \cr 
  {\bf p}_i(t + \Delta t) &=& {\bf p}_i(t) + \frac{\Delta t}{2}
  {\bf f}_i(t) \cr
  {\bf f}_i(t + \Delta t) &=& \displaystyle\sum_{j \ne i}
  {\bf F}_{ij}(t + \Delta t) \cr
  {\bf p}_i(t + \Delta t) &=& {\bf p}_i(t) + \frac{\Delta t}{2}
  \left[ {\bf f}_i(t) + {\bf f}_i(t + \Delta t) \right],
\end{eqnarray}
where $\Delta t$ is the time step size and ${\bf f}_i(t)$ is the force
acting on particle $i$ at time $t$.  The DPD units of length, mass,
and time are specific to the particular set of model parameters, and
the exact relationship between these parameters and a real fluid in a
particular situation can be determined by considering the
dimensionless groups relevant to the motion of that system, such as
the Mach, Reynolds, and Weber numbers.

Many further modifications to the model have been suggested and
implemented by others.  Some of the most interesting include energy
conservation, colloidal particles, and polymers.  In very recent work,
it has been shown that it is possible to derive a modified version of
DPD directly from the underlying molecular
dynamics~\cite{flekkoy:md,flekkoy:dpd}.

\section{Temporal Behavior of the Characteristic Domain Size}
\label{temporal}

A central quantity in the study of growth kinetics is the
characteristic domain size $R(t)$.  For binary systems in the regime
of sharp domain walls, this is usually thought to follow algebraic
growth laws of the form 
\begin{equation}\label{scaling}
  R(t) \sim t^\beta.
\end{equation}
For symmetric quenches without hydrodynamic interactions, dynamical
scaling theory and experiment~\cite{bray:ordering} indicate that the
scaling exponent $\beta = \frac{1}{3}$.  If flow effects are relevant,
typically
\begin{equation}\label{2d-exponents}
  \beta = \cases{
    \frac{1}{2} & for $R \ll{} R_h$ \qquad (diffusive)\medskip\cr
    \frac{2}{3} & for $R \gg{} R_h$ \qquad (inertial hydrodynamic)\cr
    }
\end{equation} 
in two dimensions, and
\begin{equation}\label{3d-exponents}
   \beta = \cases{
    \frac{1}{3} & for $R \ll{} R_d$ \qquad\qquad~ 
                    (diffusive)\smallskip\cr
    1           & for $R_d \ll{} R \ll{} R_h$ \qquad 
                    (viscous hydrodynamic)\smallskip\cr
    \frac{2}{3} & for $R \gg{} R_h$ \qquad\qquad~ 
                    (inertial hydrodynamic),\cr
    }
\end{equation} 
in three dimensions, where $R_h = \eta^2 /(\rho \kappa)$ is the
hydrodynamic length and $R_d = \sqrt{\eta D}$ is the diffusive length,
expressed in terms of the absolute (dynamic) viscosity $\eta$, density
$\rho$, surface tension coefficient $\kappa$, and diffusion
coefficient $D$.  These scaling laws follow directly from dimensional
analysis of the macroscopic fluid dynamics equations (so-called model
H, or Cahn--Hilliard coupled to Navier--Stokes hydrodynamics) in the
appropriate regimes~\cite{bray:ordering}.  When $R \ll{} R_d$ in three
dimensions, diffusive effects dominate and we expect the domains to
form via the Lifshitz--Slyozov--Wagner (LSW) evaporation--condensation
mechanism~\cite{bray:ordering}.  When $R \gg{} R_h$ in both two and
three dimensions, we expect the growth mechanism to be surface tension
driven by hydrodynamic flow, balanced by inertial
effects~\cite{bastea:comment}.  If $R_d \ll{} R \ll{} R_h$ in three
dimensions, we expect viscous hydrodynamic effects to dominate, as
predicted by Siggia~\cite{siggia:late}; in this regime the surface
tension drives the transport of the fluid along the interface, which
is possible only if both phases are continuous~\cite{bray:ordering,%
siggia:late}.  This third growth regime cannot occur in two
dimensions, and so we expect to see diffusive growth for $R \ll{}
R_h$~\cite{bastea:comment}.  Due to our choice of model parameters and
the small size of our simulations, we do not expect to probe the
viscous or inertial hydrodynamic regimes.

Simulations in two spatial dimensions are especially useful to
emphasize the importance of correct hydrodynamics: simulations which
do not conserve momentum typically give $\beta=\frac{1}{3}$ for the
diffusive regime ($R \ll R_h$)~\cite{cov-nov:binary,cov-nov:erratum,%
nov-cov:binary,bib:appert,emerton:microemulsions,emerton:erratum}.
Simulation methods with correct hydrodynamic interactions typically
give the expected result of $\beta=\frac{1}{2}$ (see e.g.,
Refs.~\cite{cov-nov:binary,cov-nov:erratum,nov-cov:binary,%
velasco:spinodal,velasco:off-critical,kumar:novel,bib:lwac}).  It is
worth noting that momentum conservation is not thought necessary to
model spinodal decomposition in binary metallic alloys, since phase
separation occurs by the migration of atoms to neighboring vacancies
on the crystalline lattice~\cite{fratzl:vacancy}.  Simulations based
on lattice-Boltzmann techniques also typically display
$\beta=\frac{1}{3}$.  These observations are supported by a
renormalization group approach which has shown that thermal
fluctuations cause Brownian motion-driven coalescence and play a
crucial role in causing $\beta$ to assume the value of
$\frac{1}{2}$~\cite{bray:ordering}; although some lattice-Boltzmann
techniques include these fluctuations~\cite{chen:late}, most do not.

There is some confusion in the literature over which scaling exponents
should be observed for off-critical quenches, and whether or not the
algebraic scaling law [Eq.~(\ref{scaling})] should in fact be obeyed
for any off-critical binary immiscible fluid.  Several authors have
reported the coexistence of multiple length
scales~\cite{weig:minority,wagner:breakdown,tanaka:double}.  It is
likewise uncertain as to what behavior should be observed from binary
immiscible fluids of differing viscosity~\cite{sappelt:glassy}.
Certainly, one growth mechanism we expect for off-critical quenches of
both equal and differing viscosity is the LSW
evaporation--condensation mechanism, for which
$\beta=\frac{1}{3}$~\cite{bray:ordering}.

It should also be noted that there are still some experimental and
theoretical challenges in unraveling the behavior of systems in
which both the order parameter and the momentum are locally
conserved~\cite{bray:ordering}.  Experimentally, for example, it is
difficult if not impossible to study two-dimensional fluid systems.
As far as numerical studies are concerned, it is important to
recognize that three-dimensional simulations are particularly
demanding on all the aforementioned techniques, and so definitive
results are harder to come by than in two dimensions.

Indeed, Cates's group in Edinburgh has recently reported somewhat
conflicting results from three-dimensional studies of binary
immiscible fluid separation for equal viscosity fluids using
dissipative particle dynamics and lattice-Boltzmann
methods~\cite{jury:scaling,kendon:spinodal}.  While the former
suggests the persistence of non-universal length scales, hypothesized
to be due to the intrusion of a ``molecular'' or discretization length
scale, this is not supported by the latter, from which finite-size
effects were more rigorously excluded.  Note, however, that whereas
the lattice-Boltzmann scheme was based on a Landau free-energy
approach, and is essentially no more than a finite-difference solution
of the continuum model H equations, the macroscopic equations to which
the spinodal DPD scheme corresponds have not yet been derived.  This
makes it unclear whether the two systems being simulated are really
one and the same.  Both of these studies emphasize the importance of
observing dynamical scaling over decades of time before drawing
conclusions as to the true nature of the scaling regime.  One notable
result of their work is the clarification of the relative time scales
over which computer simulation techniques typically operate.  For the
simulations they discuss, molecular dynamics time scales are on the
order of $10^2$ in dimensionless units, lattice-gas automata and
Langevin dynamics probe time scales on the order of $10^2$ through
$10^4$, DPD typically $10^3$ to $10^5$, and methods based on the
lattice-Boltzmann equation $10^1$ through $10^8$.

Grant and Elder have recently argued~\cite{grant:spinodal} that $\beta
\le \frac{1}{2}$ in any asymptotic scaling regime because the Reynolds
number $Re = \rho R V / \eta$ (where $V$ is a characteristic velocity)
cannot diverge, and in fact must remain less than a critical value
$Re_{\rm cr}$ to avoid the onset of turbulence~\cite{landau:fluid} and
possible turbulent remixing of the fluids.  The conclusion they draw
is that the $\beta=\frac{2}{3}$ scaling regime must be
transient~\cite{grant:spinodal}.  However, their analysis neglects
mention of the relative strength of the interface, quantified for
example by the Weber number $N_{\rm We} = \rho R V^2 / \kappa$.  If
$N_{\rm We}$ is small at the onset of turbulence we would expect
turbulent remixing to be delayed, or perhaps even postponed
indefinitely allowing $Re$ to diverge.  In any case, the separation
dynamics would likely be affected by $Re \ge Re_{\rm cr}$; for
example, the fluid viscosity in a turbulent region is roughly
proportional to $Re$~\cite{landau:fluid}.

\section{Estimating the Characteristic Domain Size}\label{estimating}

In order to characterize the phase separation kinetics within a binary
immiscible fluid, we need a practical tool to measure the
characteristic domain size corresponding to the state of the system at
a given point in time.  The use of the static structure function for
this purpose is widespread.  However, we have noticed bizarre ``early
time'' behavior when using the static structure function to
characterize simulations of highly off-critical quenches.
Specifically, the characteristic domain size would sometimes change
suddenly by an order of magnitude.  This abrupt change in behavior did
not correspond to anything observable in the time evolution of the
positions of the DPD particles.  Such anomalous behavior is likely due
to the large fluctuations prevalent in the static structure function
for small simulations of highly-mixed binary fluids, for which the
characteristic domain size is very small.

The radial distribution function $g(r)$ is a well-established tool for
the analysis of single-phase
fluids~\cite{allen:liquids,egelstaff:liquid}, and indicates the
likelihood of finding two particles separated by a distance $r$.  For
binary fluids we can also use the same-phase and differing-phase
distribution functions~\cite{velasco:spinodal}.  The same-phase
distribution function $g_{00}(r)$ describes the likelihood of finding
two particles of the {\em same\/} phase separated by a distance $r$,
and the differing-phase distribution function $g_{01}(r)$ describes
the likelihood of finding two particles of {\em differing\/} phase
separated by a distance $r$.  From the peaks of the differing-phase
radial distribution function, we can estimate the characteristic
separation of particles of differing phase (i.e., the characteristic
domain size).  Consequently, we decided to calculate the distribution
between particles of different phase (color),
\begin{equation}\label{gr}
  g_{01}(r) = {m\ n(r,\Delta r) \over 
    \rho N \phi [1 - \phi] V(r,\Delta r)},
\end{equation}
where $m = m_i~\forall i$ is the mass of each particle (assumed all
identical), $n(r,\Delta r)$ is the number of particle pairs of
differing phase with separation between $r - \Delta r / 2$ and $r +
\Delta r / 2$, $N$ is the total number of particles, $\phi \in (0,1)$
is the fraction of particles of one phase (the more viscous phase if
the two phases are of different viscosity), and $V(r,\Delta r)$ is the
volume of the spherical shell of radius $r$ and thickness $\Delta r$
(from $r - \Delta r / 2$ to $r + \Delta r / 2$).  It is apparent from
Eq.~(\ref{gr}) that $g_{01}(r)$ can only be calculated for $r \le L/2$
in a simulation with a periodic box size of $L$; this should not be
interpreted as an undesirable limitation as finite size effects are
normally significant for $R >
L/2$~\cite{jury:scaling,kendon:spinodal}.  A value of $\Delta r =
L/2000$ gives a reasonable compromise between noise and resolution for
the size of simulation we discuss in this paper.

The principal difficulty lies in analyzing the results automatically,
as we need to calculate $g_{01}(r)$ at many time steps within each
simulation in order to display the domain growth over time.  We chose
to calculate the median of that portion of the smoothed $g_{01}(r)$
curve extending above a threshold value $1+3s$, where $s$ is the
standard deviation of the smoothed curve when it has effectively a
constant unit value, estimated from the last tenth of the smoothed
$g_{01}(r)$ curve at $t=1$.  We used a fourth order Savitzky--Golay
smoothing filter~\cite{press:nr,bromba:smoothing} to smooth
$g_{01}(r)$ over $r$ at each point in time with a symmetric window
size chosen to reduce the noise while leaving significant features
untouched (41 points in two dimensions and 101 points in three).  We
chose the cutoff threshold of $1+3s$ so as not to bias the median by
the size of the periodic simulation box; likewise, we chose the median
in preference to the mean (first moment) because the median is less
biased by outliers, such as those 2.5\% of noisy points which
effectively have unit value but extend above the threshold.  Both the
median and mean give poor estimates of the characteristic domain size
as it approaches $L/2$, the limit of calculable $g_{01}(r)$.  When
this situation is detected, we use the global maximum of $g_{01}(r)$
to estimate the domain size instead; this is a continuous transition
for symmetrical peaks.

Once we have calculated the characteristic domain size for a series of
$g_{01}(r)$ curves taken at different times from a single simulation,
we begin a search for linear sections in the plot of $\log_{10} R$
vs.\ $\log_{10} t$.  This has also been automated, using analysis of
variance to decide whether a given section of the plot is linear or
cubic.  We used the analytic expressions for the least-squares fits
with moments taken about the means to minimize the effect of roundoff
error.  We keep only the longest linear sections, subject to there not
being any significant gaps in the coverage of the log--log plot.  An
ensemble average of a number of simulations yields a plot of the
scaling exponent $\beta$ vs.\ $\log_{10} t$, in which long horizontal
(zero gradient) sections represent algebraic growth.  This procedure
allows more accurate determination of the scaling exponent than a
visual comparison of $\log_{10} R$ vs.\ $\log_{10} t$ to a straight
line of a particular slope, and provides a statistically valid
determination of whether or not the growth is truly algebraic over a
particular time period.  Finite size effects for $R \approx L/2$
normally result in non-algebraic growth or unusually low scaling
exponents; either case is easily detected by our method.  A further
advantage is the automation we have described, permitting large
ensemble averages with minimal effort.

We should note that estimating the characteristic domain size using
the median can occasionally lead to discontinuities.  If these
discontinuities are large enough they will cause a break between
linear sections.  Small discontinuities may be spanned by a single
line, which would affect the slope ($\beta$) and be detrimental to the
results.  However, large discontinuities will not be spanned, and so
will not affect the slope.

The results we obtained using these techniques are considerably better
than those obtained with the static structure function, especially for
highly-mixed fluids where the pair distribution function is not
drowned by fluctuations to the extent that the static structure
function is.  Furthermore, the pair distribution function is more
intuitive to analyze than the static structure function, which aids
our interpretation of the results.

\section{Simulation Results}\label{results}

For the simulations of fluids with differing viscosity, we chose one
phase to be an order of magnitude more viscous than the other, i.e.,
$\gamma_1 = 10 \gamma_0$.  Before simulating this new complex fluid,
it is advisable to verify that increasing the parameter $\gamma$,
while keeping the temperature constant, does indeed increase the
viscosity.  For the lower-viscosity fluid we chose the model
parameters shown in Table~\ref{LowerViscosityParameters}.  The
absolute (dynamic) viscosity of a homogeneous DPD fluid can be
estimated theoretically from the continuous-time viscosity, ignoring
the effect of the conservative forces (i.e.,
$\alpha$=0)~\cite{marsh:properties,marsh:thesis}; the continuous-time
viscosity of this lower-viscosity fluid is $\eta = 2.8$.

In order to verify this estimate, we performed a series of simulations
of steady shear using Lees--Edwards periodic boundary conditions.  We
performed a total of 63 simulations with a time step of $\Delta t =
0.1$ (in our DPD time units), each from a different random initial
configuration and each allowed to settle to steady-state shear before
measurement began.  We studied systems of both 1600 and 6400
particles, six simulations at each of nine different shear rates for
the former and three simulations at each of three distinct shear rates
for the latter.  As the results from the larger simulations gave a
mean viscosity nearly identical to that of the smaller ones, we can
conclude that finite size effects do not bias the viscosity of the
smaller, faster simulations.  We calculated the velocity profile for
each set of parameters, and found it to be statistically
indistinguishable from linear in every case.

Analyzing these simulations led to a conclusion of $\eta = 1.94 \pm
0.01$.  Others have also found discrepancies between theory and
simulation, particularly regarding the kinematic contribution to
viscosity~\cite{marsh:properties}, so the difference between the
simulated viscosity and the continuous-time viscosity is not entirely
surprising.  Since molecular dynamics simulations containing only
conservative forces give a finite viscosity, we would be surprised if
the theoretical estimate did not differ from our calculations.

In order to measure the viscosity of the more viscous fluid, we set up
a series of 1600 particle simulations of a homogeneous DPD fluid as
described above, using a total of 30 simulations.  We varied the shear
rate in the range that gave the best results in the previous
simulations.  The model parameters differed in that $\gamma$ was a
factor of ten larger while $\sigma$ was a factor of $\sqrt{10}$ larger
to keep the temperature constant [see Eq.~(\ref{temperature2})]; these
parameters are shown in Table~\ref{HigherViscosityParameters}.  We
decreased the time step to $\Delta t = 0.01$ due to the increased
magnitude of the dissipative and stochastic forces.  Analyzing these
results led to the conclusion that $\eta = 17.2 \pm 0.3$, which
confirms the increase in viscosity.  This increase is close to a
factor of ten, so that for similar model parameters it is reasonable
to conclude that $\gamma$ is approximately proportional to viscosity.

It is possible to measure the surface tension of a binary immiscible
fluid by integrating the pressure tensor across a flat interface, or
by verifying Laplace's law for a series of equilibrium bubbles of
varying radii~\cite{allen:liquids}.  Laplace's law was verified for a
DPD binary immiscible fluid in our previous
simulations~\cite{cov-nov:binary,cov-nov:erratum,nov-cov:binary}, and
so surface tension measurements were omitted from the present study.
These calculations would, however, allow identification of the unit of
time for comparison with other simulation techniques (see e.g.,
Refs.~\cite{jury:scaling,kendon:spinodal}).  Theoretical estimates for
surface tension are also available, but are of similar accuracy to the
viscosity estimate above.

For the simulation of fluids of differing viscosity, we must also
choose the relatively small time step size of $\Delta t = 0.01$ in
order to ensure the stability of the algorithm as a result of the
increased magnitude of the dissipative forces.  This has the
consequence of making these simulations computationally much more
expensive than equal-viscosity simulations.  In this context, it is
worth commenting on the virtues of using DPD to perform simulations of
differing-viscosity fluids compared with other models.  Sappelt and
J{\"a}ckle use an approach based on the Cahn--Hilliard equation
(so-called model B without noise) with a concentration-dependant
mobility~\cite{sappelt:glassy} so do not include hydrodynamics, unlike
our approach with DPD.  The other mesoscale techniques which could be
used to model these fluids include lattice-gas automata and methods
based on the lattice-Boltzmann equation.  As with DPD, there is a high
computational price to the lattice-gas approach, which requires
adjustment to the collisional outcomes of the look-up tables.
Lattice-Boltzmann (more correctly, lattice-BGK) models can be used,
based for example on the Swift--Osborn--Yeomans free energy functional
approach~\cite{swift:nonideal}, but there is a similar computational
price, although with the additional complication of poorly understood
numerical instabilities.  However, DPD offers the simplest algorithmic
implementation which is thermodynamically consistent.

Our differing-viscosity simulations used the model parameters shown in
Table~\ref{DifferingModelParameters}.  Each simulation had 6400
particles, and ran for 50,000 time steps.  We chose to use only 6400
particles in our simulations so that individual simulations would be
quick enough to be run multiple times, allowing us to consider the
effect of changing the phase fraction and viscosity, and to increase
the confidence in our results and calculate accurate error estimates
with ensemble averaging.  Simulations with more particles would have
given better resolution of small scale features (relative to the
system size) and postponed the finite size effects, at the price of
increased computational demands.  Our resources would only have
allowed a few simulations of the size and length used by Jury {\em et
al.\/}~\cite{jury:scaling} (1,000,000 particles), making studies of
the sort we describe in this paper impossible.

\begin{figure}\noindent
  \includegraphics{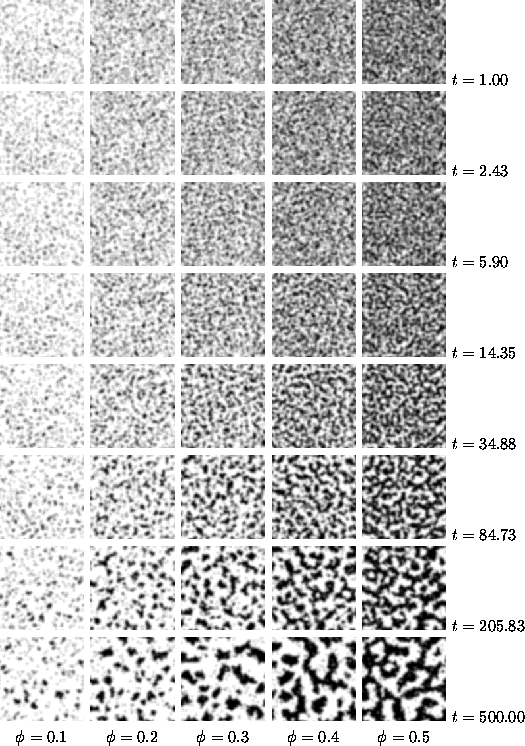}
  \vspace{1cm}
  \caption{Time-evolution of five sample simulations of
    two-dimensional differing-viscosity fluids, each simulation having
    a different viscous phase fraction (varying from $\phi=0.1$
    through $\phi=0.5$).}\label{2d-glassy-time}
\end{figure}

\begin{figure}\noindent
  \includegraphics{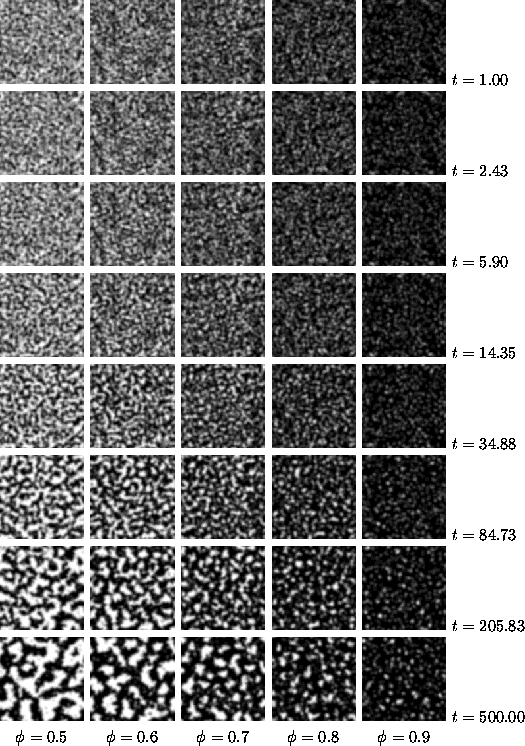}
  \vspace{1cm}
  \caption{Time-evolution of five sample simulations of
    two-dimensional differing-viscosity fluids, each simulation having
    a different viscous phase fraction (varying from $\phi=0.5$
    through $\phi=0.9$).}\label{2d-glassy-time2}
\end{figure}

\begin{figure}\noindent
  \includegraphics{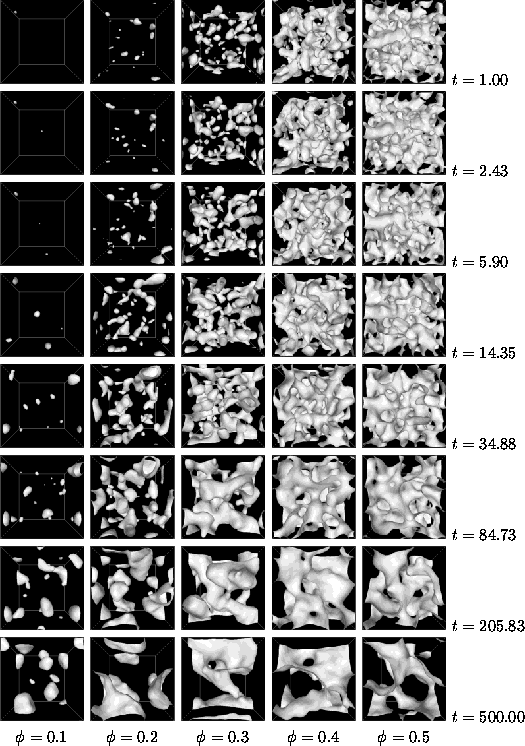}
  \vspace{1cm}
  \caption{Time-evolution of five sample simulations of
    three-dimensional differing-viscosity fluids, each simulation
    having a different viscous phase fraction (varying from $\phi=0.1$
    through $\phi=0.5$).}\label{3d-glassy-time}
\end{figure}

\begin{figure}\noindent
  \includegraphics{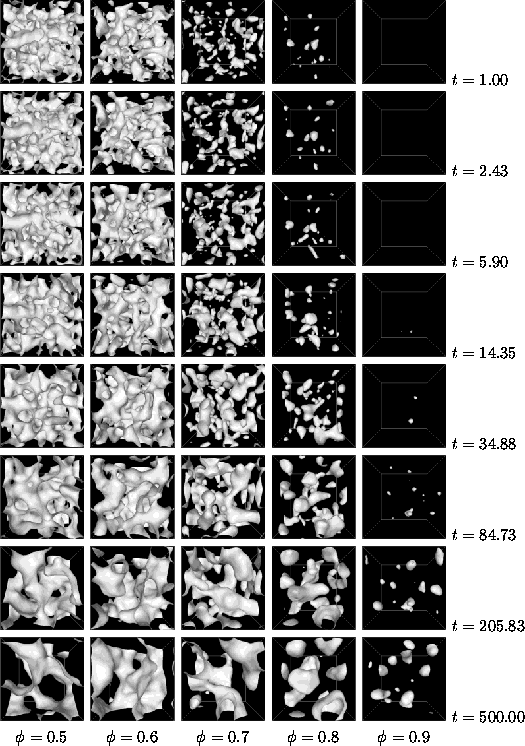}
  \vspace{1cm}
  \caption{Time-evolution of five sample simulations of
    three-dimensional differing-viscosity fluids, each simulation
    having a different viscous phase fraction (varying from $\phi=0.5$
    through $\phi=0.9$).}\label{3d-glassy-time2}
\end{figure}

\begin{figure}
  \includegraphics{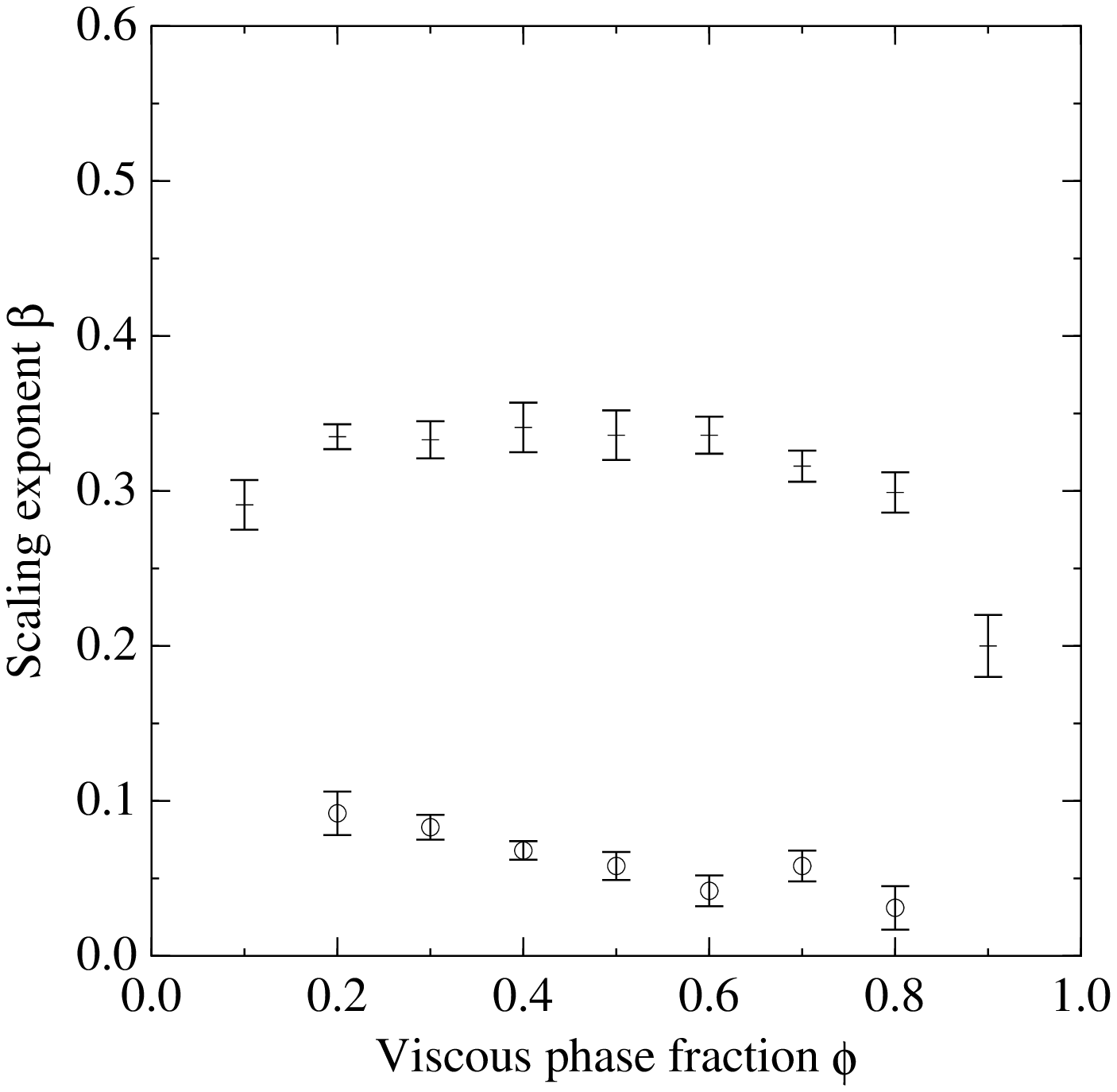}
  \vspace{1cm}
  \caption{Scaling exponents for two-dimensional differing-viscosity
    fluids as a function of viscous phase fraction.  Circles indicate
    ``early time'' and horizontal marks ``late time''; error bars are
    68\% confidence intervals.}\label{glassy2}
\end{figure}

\begin{figure}
  \includegraphics{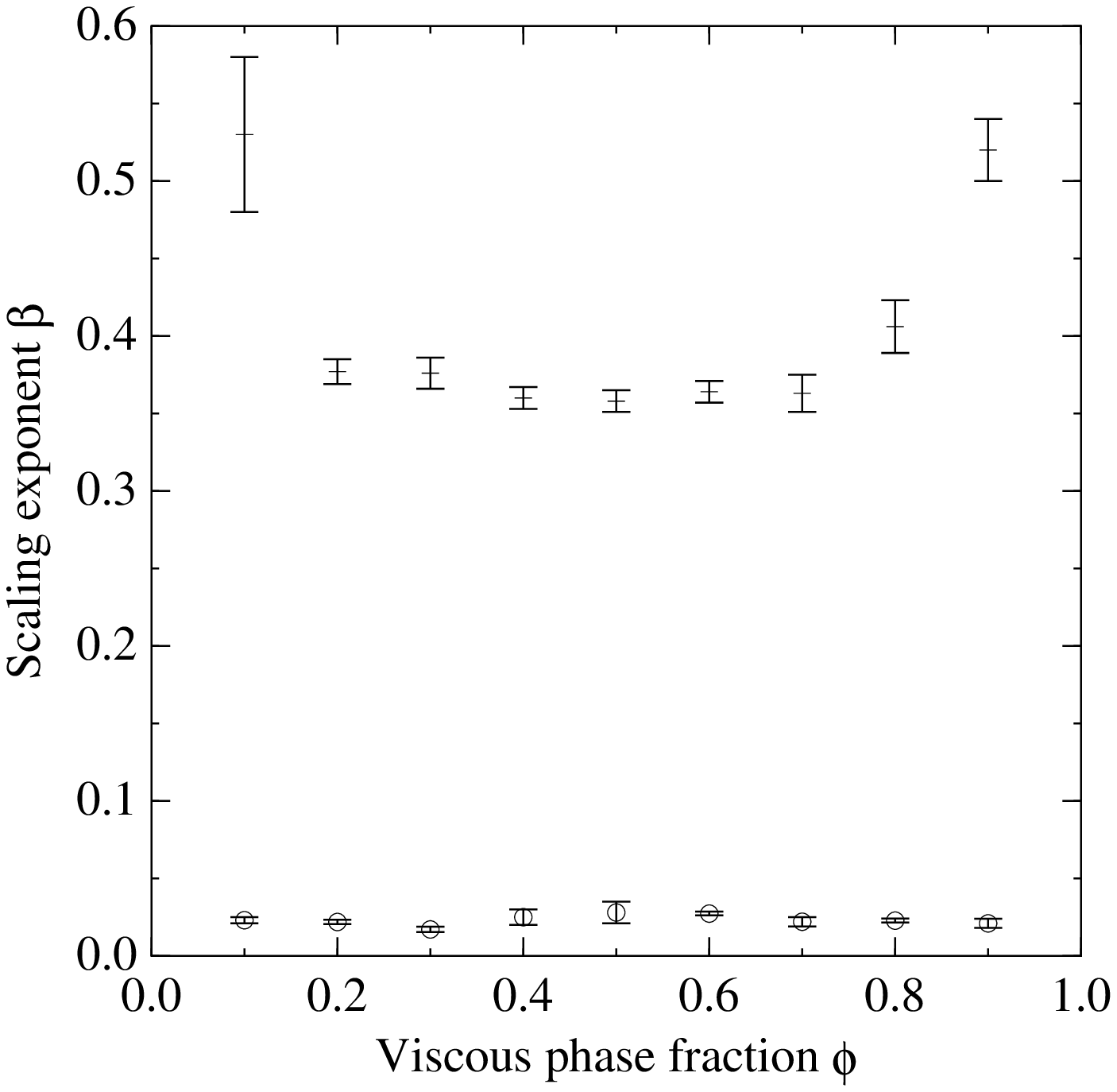}
  \vspace{1cm}
  \caption{Scaling exponents for three-dimensional differing-viscosity
    fluids as a function of viscous phase fraction.  Circles indicate
    ``early time'' and horizontal marks ``late time''; error bars are
    68\% confidence intervals.}\label{3d-glassy2}
\end{figure}

We calculated the pair distribution function at 64
logarithmically-spaced points in time during the course of each
simulation, starting from time $t=1$ and finishing at $t=500$ (in our
DPD time units).  In both two and three dimensions, we performed ten
simulations at each of nine different fractions of the viscous phase,
ranging from $\phi = 0.1$ (10\% viscous phase) to $\phi = 0.9$ (90\%
viscous phase).  We show the time evolution of a single simulation at
each value of the viscous fraction in
Figs.~\ref{2d-glassy-time}--\ref{3d-glassy-time2} for both two and
three dimensions.  We represent the positions of the DPD particles in
two dimensions by a gray scale map, in which the particle positions
are weighted by Eq.~(\ref{omega}) and assigned an intensity of gray
based upon the proportion of each phase in this localized-average.
Figures~\ref{3d-glassy-time} and~\ref{3d-glassy-time2} display the
three-dimensional surface of the interface between the two immiscible
phases, as defined by there being equal proportions of each fluid in
the localized-average.  In these four figures the gradual development
of domains can be seen.  We examined the results of the domain size
analysis for each simulation in plots of $\log_{10} R$ vs.\ $\log_{10}
t$, and further examined each ensemble average in plots of $\beta$
vs.\ $\log_{10} t$; the latter yielded the mean values of $\beta$ and
corresponding 68\% confidence intervals for significant sections of
algebraic growth.  We show these results for two and three dimensions
in Tables~\ref{glassy} and~\ref{3d-glassy}, and Figs.~\ref{glassy2}
and~\ref{3d-glassy2}.  The range of $\log_{10} t$ we give in the
tables should be taken as a rough guide, since there is often a high
degree of correlation between the start and end times of a particular
linear section and its growth exponent within the broad category of
late time.  We should emphasize that we are using the term ``late
time'' as a relative category in this paper, to distinguish these
results from those obtained at very early times.  Due to the model
parameters and small size of these simulations, we do not expect to
probe the viscous or inertial hydrodynamic regimes (see
Eqs.~(\ref{scaling})--(\ref{3d-exponents}), and compare with
Refs.~\cite{cov-nov:binary,cov-nov:erratum,nov-cov:binary,%
novik:finite-difference}).

\begin{figure}\noindent
  \includegraphics{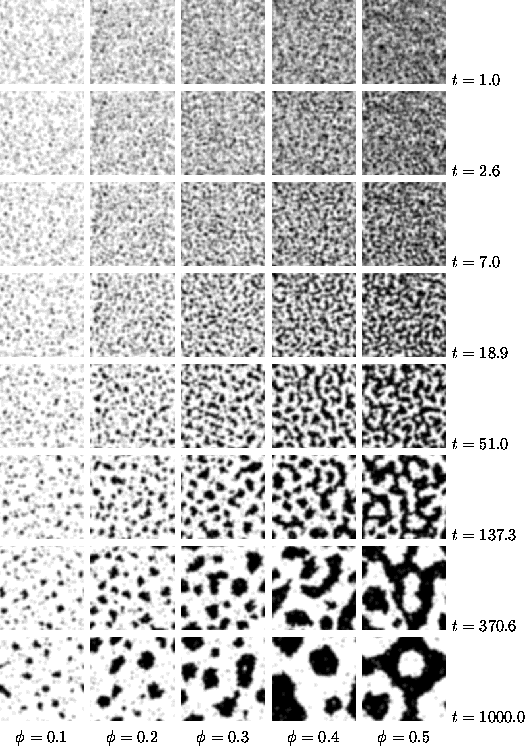}
  \vspace{1cm}
  \caption{Time-evolution of five sample simulations of
    two-dimensional equal-viscosity fluids, each simulation having a
    different minority phase fraction (varying from $\phi=0.1$
    through $\phi=0.5$).}\label{2d-same-time}
\end{figure}

\begin{figure}\noindent
  \includegraphics{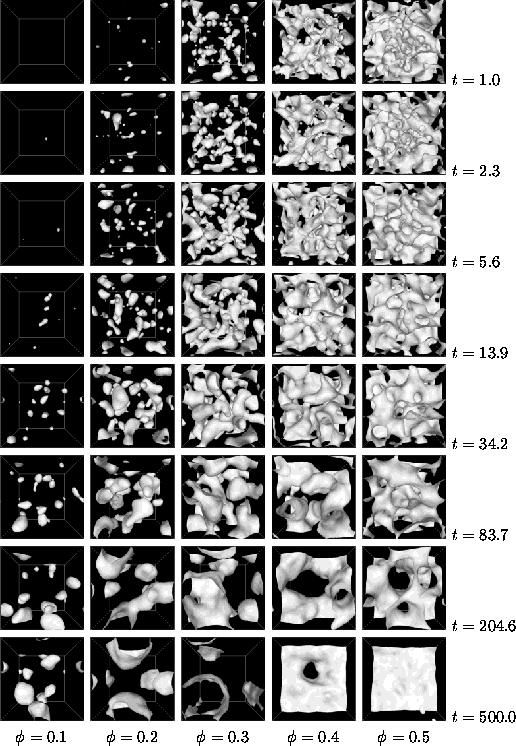}
  \vspace{1cm}
  \caption{Time-evolution of five sample simulations of
    three-dimensional equal-viscosity fluids, each simulation having a
    different minority phase fraction (varying from $\phi=0.1$
    through $\phi=0.5$).}\label{3d-same-time}
\end{figure}

\begin{figure}
  \includegraphics{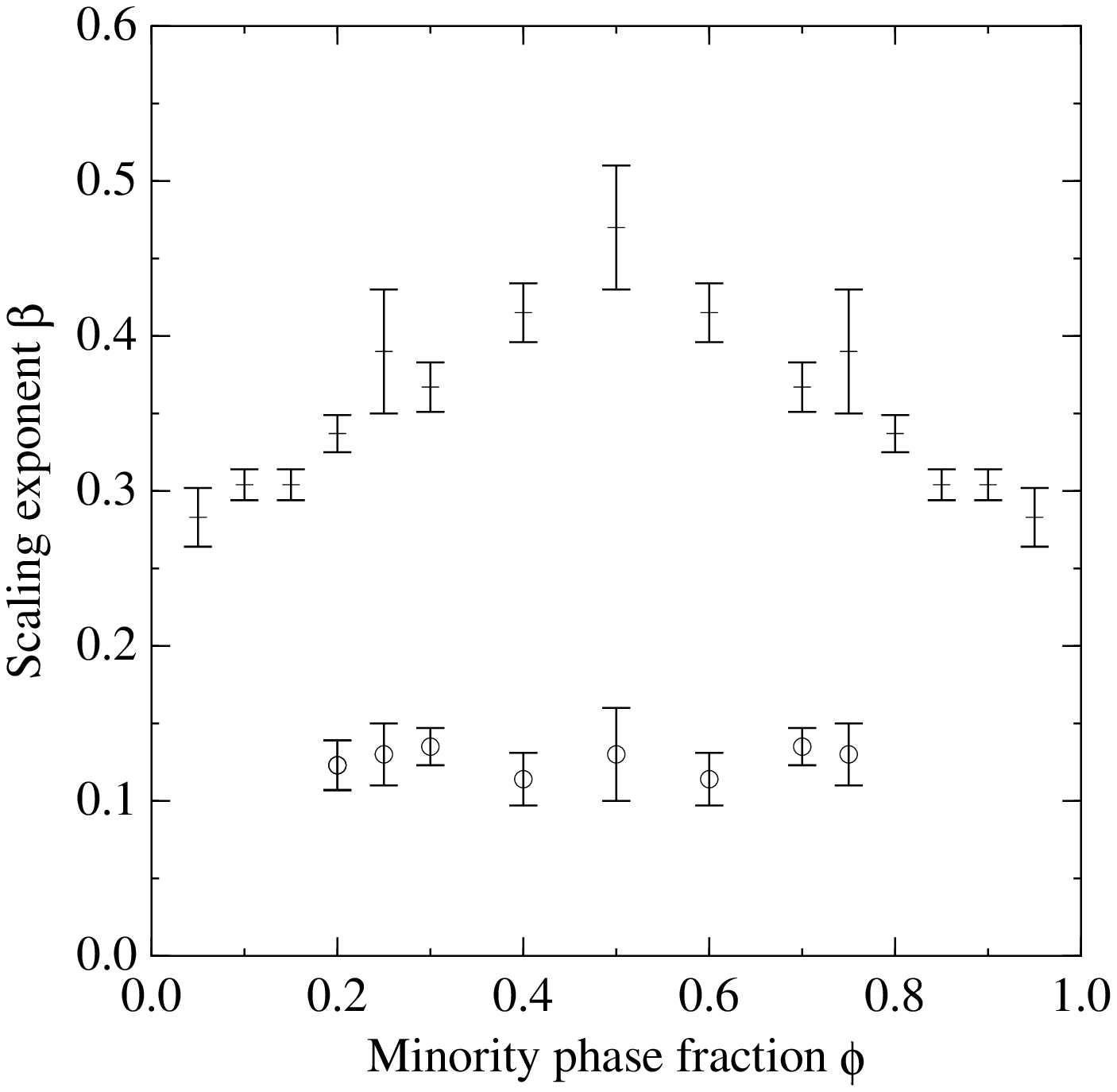}
  \vspace{1cm}
  \caption{Scaling exponents for two-dimensional equal-viscosity
    fluids as a function of minority phase fraction.  Circles indicate
    ``early time'' and horizontal marks ``late time''; error bars are
    68\% confidence intervals.}\label{same-viscosity2}
\end{figure}

\begin{figure}
  \includegraphics{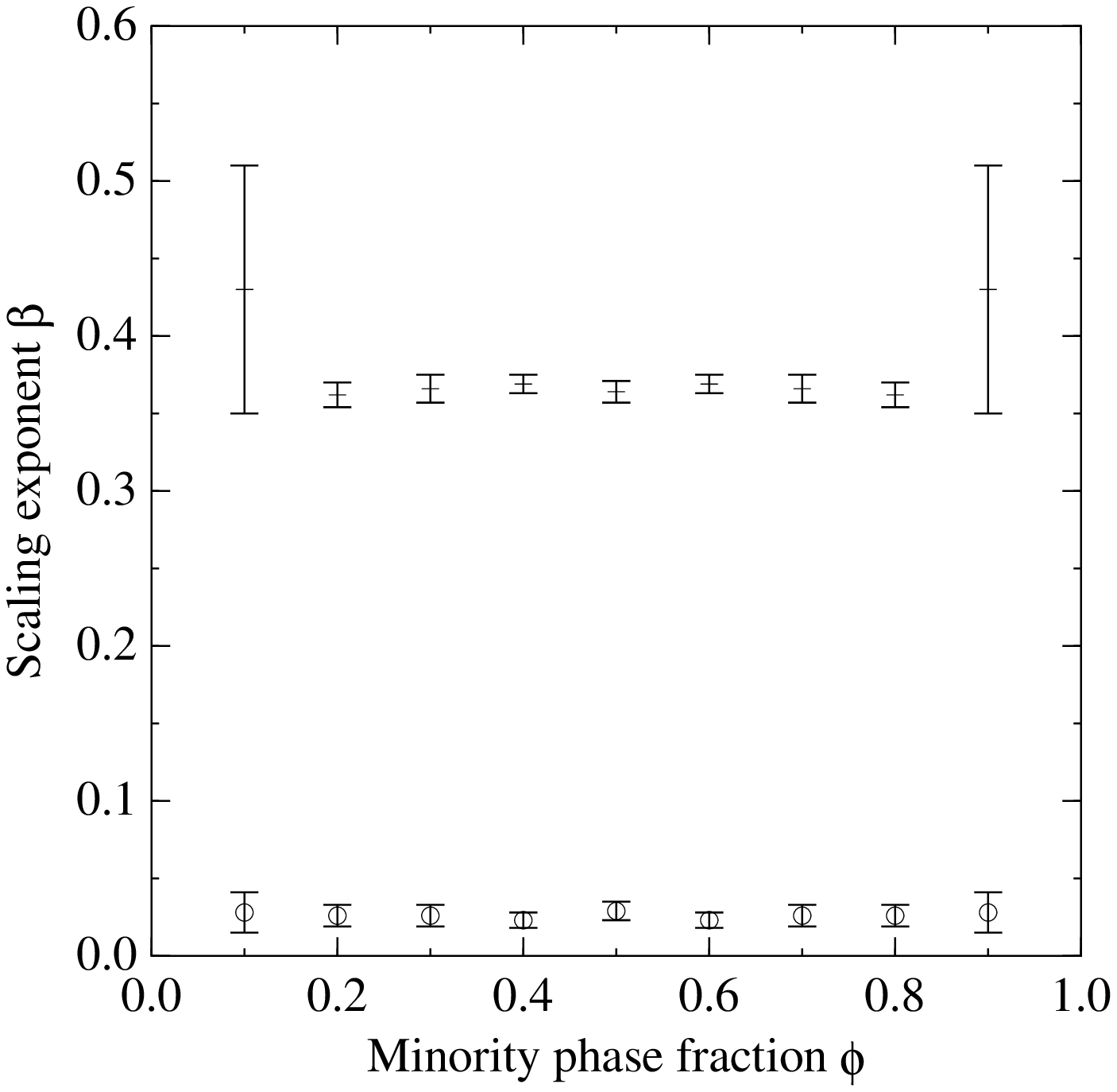}
  \vspace{1cm}
  \caption{Scaling exponents for three-dimensional equal-viscosity
    fluids as a function of minority phase fraction.  Circles indicate
    ``early time'' and horizontal marks ``late time''; error bars are
    68\% confidence intervals.}\label{3d-same2}
\end{figure}

We also constructed a series of equal-viscosity simulations in both
two and three dimensions.  These had $\Delta t = 0.1$, 6400 particles,
and stopped at $t=1000$.  The model parameters were the same as in the
differing-viscosity simulations, with the exception that both phases
were identical to the less-viscous phase of the differing-viscosity
simulations (i.e., $\gamma = \gamma_0$ and $\sigma = \sigma_0$).  We
calculated the pair distribution function at 64 different points from
$t=1$ to $t=1000$, with ten simulations at each of eight different
minority phase fractions from $\phi = 0.05$ to $\phi = 0.5$ in two
dimensions, and ten simulations at each of five different fractions
from $\phi = 0.1$ to $\phi = 0.5$ in three dimensions.  We show the
time evolution of a single simulation at each value of the minority
fraction in Figs.~\ref{2d-same-time} and~\ref{3d-same-time} for two
and three dimensions respectively.  In these two figures the gradual
development of domains can be seen, and is greatly slowed for small
minority fraction.  We show the scaling exponents for two and three
dimensional equal-viscosity fluids in Tables~\ref{same-viscosity}
and~\ref{3d-same}, and Figs.~\ref{same-viscosity2} and~\ref{3d-same2},
where we mirror the scaling exponents about $\phi=0.5$ to show the
full range of minority phase fraction.  Comparison with
Figs.~\ref{2d-same-time} and~\ref{3d-same-time} qualitatively confirms
the same behavior.

\section{Discussion}\label{discussion}

\begin{figure}
  \includegraphics{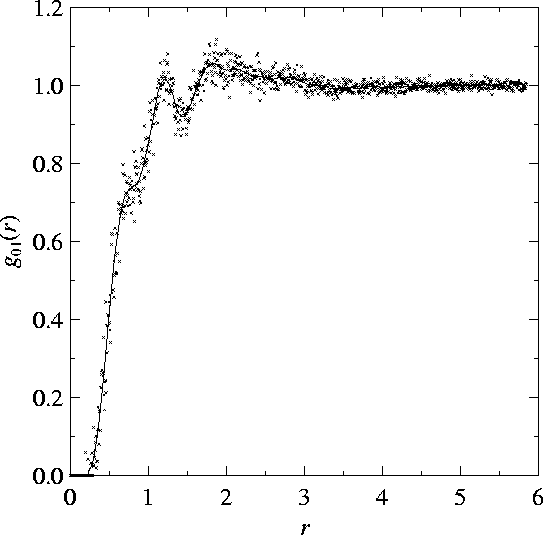}
  \vspace{1cm}
  \caption{Differing-phase radial distribution function for a
    three-dimensional equal-viscosity fluid ($\phi=0.5$) at
    $t=13.9$.  The unit of length for $r$ is specific to the
    particular set of model parameters used.}\label{g00013.9}
\end{figure}

A feature common to all these simulations is the lack of
scale-invariance at very early times, until approximately $t=10$ to
$t=50$ (in our DPD time units) depending on the exact composition of
the fluid.  This is apparent in the $g_{01}(r)$ curves as multiple
peaks of similar magnitude, but could not be seen in the frequency
domain shown by the static structure function.  Figure~\ref{g00013.9}
shows an example of the differing-phase radial distribution function
for a three-dimensional equal-viscosity simulation ($\phi=0.5$) at
$t=13.9$; the crosses represent the actual data and the curve
represents the smoothed data.  This is one of the simulations we show
in Fig.~\ref{3d-same-time}.  The multiple peaks typically evolve by
changing their height and relative weight, but not their position.
This is noticeably different from our ``late time'' behavior of the
distribution function, where a single peak gradually advances its
position while broadening but retaining its height and general shape.
It is because of this ``early time'' behavior that it was decided not
to use solely the global maximum of $g_{01}(r)$ to estimate the
characteristic domain size.

At these ``early times'' in all of the simulations, we observed
algebraic growth with a very small exponent, roughly $0.126 \pm 0.003$
in two dimensional equal-viscosity simulations, $0.062 \pm 0.009$ in
two dimensional differing-viscosity simulations (except $\phi=0.9$
(90\% viscous phase), for which $\beta=-0.24 \pm 0.07$), and $0.0246
\pm 0.0008$ for simulations in three dimensions.  This corresponds to
the region of breakdown of scale invariance we described at the
beginning of this section.  These ``early time'' exponents are
unaffected by viscous or minority phase fraction in all of the
simulations, except the differing-viscosity simulations in two
dimensions.  Here $\beta$ decreases with increasing $\phi$ (viscous
phase fraction), probably due to the domain-growth arresting effect of
increasing viscosity.  We observed a remarkable (though short) regime
of $\beta=-0.24 \pm 0.07$ for $\phi=0.9$ (90\% viscous phase), for
which we have no adequate explanation.  Others normally discard
similar early time regimes without comment~\cite{laradji:spinodal,%
bib:lwac,chen:late,osborn:lb,gonnella:lamellar,gonnella:droplet,%
sappelt:glassy} or as an ``early stage'' or ``transient''
regime~\cite{jury:scaling,chen:late}.  However, there is growing
evidence for the coexistence of multiple domain sizes and hence a
breakdown in universality, at least in certain phase-ordering
domains~\cite{weig:minority,wagner:breakdown,tanaka:double}.

For ``late time'' domain growth in the two-dimensional
differing-viscosity simulations (see Fig.~\ref{glassy2}), we observed
a fairly constant value of $\beta$ for $\phi=0.2$ (20\% viscous phase)
through $\phi=0.6$ (60\% viscous phase), decreasing both for
$\phi=0.1$ and very slightly for $\phi=0.7$ and $\phi=0.8$, then
decreasing sharply at $\phi=0.9$.  This asymmetry is consistent with
the variation of the ``early time'' exponent, in that an increasingly
viscous fluid is expected to develop domains more gradually.  At
increasingly rarefied fractions ($\phi=0.1$ and $\phi=0.9$), domain
growth is retarded by the increased isolation of the droplets.  The
``late time'' growth exponent throughout is effectively $\frac{1}{3}$,
which suggests that the presence of fluids of differing viscosity
interferes with the normal $\beta=\frac{1}{2}$ growth mechanism in two
dimensions.  The growth exponent of $\frac{1}{3}$ is expected from the
LSW evaporation--condensation mechanism~\cite{bray:ordering}.  This is
in some ways analogous to the effect obtained by deliberately breaking
momentum conservation in symmetric quenches, as described
previously~\cite{cov-nov:binary,cov-nov:erratum,nov-cov:binary}.  Our
domains are considerably less circular at all viscous fractions than
those observed by Sappelt and J{\"a}ckle (compare
Figs.~\ref{2d-glassy-time} and~\ref{2d-glassy-time2} with
Ref.~\cite{sappelt:glassy}).  This is likely due to the lack of
hydrodynamic interactions in their model and to the greater difference
in viscosity between their two phases.  As in their simulations, our
two-phase structure for fluids of differing viscosity is not very
different from the structure for fluids of equal viscosity.  Moreover,
our simulations do not reveal any new insights regarding interfacial
structure.

In three dimensions, the ``late time'' domain growth of
differing-viscosity fluids displays nearly the opposite behavior, with
the scaling exponent increasing as the viscous fraction reaches its
extremes.  This could be explained by the increased fluid mobility in
simulations with an extra spatial dimension, as the majority phase is
completely connected and so the domain growth could occur according to
the $\beta=\frac{2}{3}$ mechanism, which is surface tension driven by
hydrodynamic flow, balanced by inertial effects.  This is
qualitatively substantiated by inspection of
Figs.~\ref{3d-glassy-time} and~\ref{3d-glassy-time2}, where a larger
degree of connectivity of the majority phase can be observed at the
extremes of viscous fraction than for $\phi \approx 0.5$.  However, a
more obvious mechanism for the domain growth would be the
$\beta=\frac{1}{3}$ LSW evaporation--condensation
mechanism~\cite{bray:ordering}; it may be a combination of these two
mechanisms that leads to our observed $\frac{1}{3} < \beta <
\frac{2}{3}$ growth.  A slight asymmetry in the ``late time'' growth
exponent is also evident, with domain growth proceeding more slowly
with increasingly viscous fluids of non-extreme viscous fraction.

For equal-viscosity fluids in two dimensions we observed the expected
$\beta=\frac{1}{2}$ for symmetric quenches
($\phi=0.5$)~\cite{cov-nov:binary,cov-nov:erratum,nov-cov:binary}.  As
we reduced the minority phase fraction, we observed a steady decrease
in the scaling exponent until $\beta=\frac{1}{3}$ is reached at the
extremes.  This confirms the results of other
workers~\cite{weig:minority} that increasingly off-critical quenches
retard the domain growth, while providing support for the observed
slowdown of growth at the extremes of viscous fraction in
differing-viscosity fluids in two dimensions, which we commented on
above.

In three dimensions, the domain growth of equal-viscosity fluids
appears largely unaffected by varying the minority phase fraction.
Although there may be some increase in $\beta$ at the extremes of
$\phi$ (as seen in the differing-viscosity fluid in three dimensions),
this is difficult to confirm definitely because of the large variation
in rate of growth observed for the simulations with $\phi=0.1$, and
hence correspondingly large confidence interval.  The scaling exponent
throughout is close to $\beta=\frac{1}{3}$.  Whereas Jury {\em et
al.\/}~\cite{jury:scaling} were intending to probe the viscous or
inertial hydrodynamic regimes with their DPD simulations of
equal-viscosity fluids in three dimensions, we aimed only to probe
length scales below $R_d$ and $R_h$ [see
Eqs.~(\ref{scaling})--(\ref{3d-exponents})].  As such, our results are
fully consistent with theirs.  Our exclusion of finite size effects is
more rigorous than theirs, and although not as extreme as that
advocated by Kendon {\em et al.\/}~\cite{kendon:spinodal} our method
gives statistical confidence that these domains are scaling
algebraically.  Both Jury {\em et al.\/}~\cite{jury:scaling} and
Kendon {\em et al.\/}~\cite{kendon:spinodal} were able to cover the
time domain more fully in three-dimensional equal-viscosity symmetric
quenches only at the cost of performing a large number of
computationally very intensive and very expensive simulations.

\section{Conclusions}\label{conclusions}

In this paper, we have described simulations of the domain growth and
phase separation of hydrodynamically-correct binary immiscible fluids
of differing and equal viscosity as a function of minority phase
concentration in both two and three spatial dimensions.  Due to our
choice of model parameters and the small size of our simulations, we
did not expect to probe the viscous or inertial hydrodynamic regimes.
In three dimensions, we found that the characteristic domain size
scales as $t^{1/3}$ for simulations of differing and equal-viscosity
fluids developing from symmetric and slightly off-critical quenches.
For highly off-critical quenches we observe an increase in the scaling
exponent.  In two dimensions, we also observe $t^{1/3}$ in simulations
of differing-viscosity fluids developing from symmetric and slightly
off-critical quenches, although we observe a decrease in the scaling
exponent for highly off-critical quenches.  In equal-viscosity fluids
in two dimensions, we observe $t^{1/2}$ for symmetric quenches and a
roughly linear decrease to $t^{1/3}$ for highly off-critical quenches;
these results are in agreement with similar lattice-gas simulations in
two dimensions~\cite{weig:minority}.

Obtaining meaningful results for ensemble averages of highly
off-critical binary immiscible fluids was only made feasible by our
automation of the calculation of the characteristic domain size by the
pair correlation function.  It also made possible the identification
of a regime of breakdown of scale invariance at very early times,
which was not noticeable in our original analysis using the static
structure function.  Further simulations aimed to probe the viscous
and inertial hydrodynamic regimes [see
Eqs.~(\ref{scaling})--(\ref{3d-exponents})] would be a useful addition
to this work, as would simulations aimed to cover longer periods of
time; however, both would require substantially increased
computational work.

\section*{Acknowledgments}

Many helpful discussions were had with Peter Bladon, Bruce Boghosian,
Alan Bray, Mike Cates, Pep Espa{\~n}ol, Simon Jury, Colin Marsh, John
Melrose, and Julia Yeomans; extra thanks is due to Matthew Probert and
Matt Segall.  We would also like to thank NATO for a grant which has
supported this work in part, the EPSRC (U.K.) E7 Grand Challenge in
Colloidal Hydrodynamics for providing access to the Cray T3D at the
Edinburgh Parallel Computing Centre, the CSAR service at the
University of Manchester for access to their Cray T3E and SGI
Origin2000 through the EPSRC High Performance Computing for Complex
Fluids grant, and the Cambridge High Performance Computing Facility
for access to their Hitachi SR2201.  KEN gratefully acknowledges
financial support from NSERC (Canada) and the ORS Awards Scheme
(U.K.).

\appendix
\section*{High-Performance Computing}

In this appendix we provide a few comments regarding the running of
DPD simulations on high performance computers.  We usually had easy
access to single-processor workstations, with a large variety of types
and speeds of processors.  Multi-processor machines allowing parallel
execution of simulations are much less common and are more difficult
to obtain access to, although they have become more common during this
research project.  We used both the Cray T3D of the Edinburgh Parallel
Computing Centre (EPCC) and the Hitachi SR2201 of the Cambridge High
Performance Computing Facility (HPCF) for computing the results
described in this paper; the former consisted of 512 processor nodes
and the latter consists of 256 nodes.

The implementation of the dissipative particle dynamics algorithm is
very similar to that of conventional molecular dynamics
algorithms~\cite{allen:liquids}.  For example, we divide the periodic
spatial domain (the simulation cell) into a regular array of equally
sized link-cells, such that each side of the rectangular domain has an
integer number of cells and each cell is at least $r_c$ across.  Each
link-cell consists of a dynamically allocated array of particles, and
pointers to the neighboring cells.  Individual particles consist of
the position--momentum vector pair and a color index.

For each time step we iterate through the particles in each link-cell,
calculating the force acting on each particle as it interacts with the
particles in the same and neighboring link-cells.  Since the DPD force
acts between pairs of particles, we must ignore half of the
neighboring cells to avoid duplication.  When considering a different
particle pair, we compare the square of the separation distance with
$r_c^2$, skipping to the next particles if the pair is out of range.
We then compute the new position and velocity as determined by the
finite-difference algorithm (see Eq.~(\ref{groot-warren}) and
Ref.~\cite{novik:finite-difference}).

We may write the complete state of the system to file, and we can
perform other calculations thereafter, for example to determine the
temperature and pressure of the system.  We used the freely-available
{\em Gnu-make\/} utility to dictate the compilation process, since the
decision structures it contains make it simple to write programs
portable to a large range of architectures.  We created a
comprehensive, automated test suite to make it easy to verify that
optimizations of the calculations did not accidentally change the
results of the computations.

Given constant $r_c$ and number density $n = \rho / m$, the DPD
algorithm scales linearly (in both computation time and memory size)
with increasing number of particles ($N$), and is limited by
computation time on all but the smallest machines.  The main
simulations we performed for this paper consisted of 6400 particles,
and it is on the parallel and serial performance of this size of
simulation that we will make most of the following comments.  Details
of the performance of this size of simulation in two dimensions are
shown in Tables~\ref{hpc-small1} and~\ref{hpc-small2}.  These tables
give the elapsed time per node in seconds and relative parallel
efficiency for the first 1000 time steps, including data for a variety
of computers and partition sizes.  These data are for code compiled
with the highest level of optimization, including some small
reductions in floating-point accuracy.  Table~\ref{hpc-small1}
describes the computers used to calculate the results in this paper,
while Table~\ref{hpc-small2} describes the computers to which we have
recently been allowed access, such as the Computer Services for
Academic Research (CSAR) Cray T3E and SGI Origin2000 in Manchester.

A typical simulation of 50,000 time steps takes 2.5~hours on a 350~MHz
Intel Pentium~II PC, the fastest single-processor machine to which we
had common access.  This same simulation would take 2.9~hours on a
16-node partition of the T3D at the EPCC\@.  However, to minimize
fragmentation of the machine, jobs using up to 32 nodes were limited
to a total execution time of 30~minutes.  One possibility was to break
up the run into 30~minute portions, but this introduces additional
overhead and complications; however, new jobs start instantly because
they need not be queued.  A better option was to run jobs on a 64-node
partition, task farming four 16-node jobs to run simultaneously.
There was a 12~hour limit to 64--512 node jobs (6~hours during the
week), but there was often a long wait in the queues.  If the
efficient usage of billed time was a significant concern, sixteen
2-node jobs would complete in 8.0~hours.  However, during the week
this meant restarting halfway through and waiting in the queue again.
Similar comments apply to the Hitachi SR2201, although its queues were
limited to 8~hours maximum.  The extra administrative overheads
involved and the billed usage means that we usually concentrated
computation on the serial workstations.  However, parallel execution
becomes more attractive with larger simulations.

The parallel efficiency of DPD with 6400 particles is good only for a
modest number of processor nodes.  This is particularly true of the
more modern parallel machines such as the Cray T3E and SGI Origin2000,
which are proportionally faster in processing than communicating when
compared with their older counterparts.  Much better parallel
efficiency has been observed with larger simulations.  The Cray T3D
shows an unusual increase in efficiency when going from a serial
calculation to a 2-node parallel calculation with 6400 particles; this
could be explained by any number of hardware-specific arguments.  We
should note that the results for the Origin2000 include the effect of
sharing the machine with other users, unlike all the other machines
whose results appear in Tables~\ref{hpc-small1} and~\ref{hpc-small2},
for which each node was dedicated to our calculations.

We decided to write the main simulation program in C/C++ as opposed to
Fortran.  This choice was made because C/C++ were believed to be the
most appropriate languages for dealing with DPD simulations which
consist of a large amount of book-keeping wrapped around fairly simple
computations.  C/C++ and Fortran are highly portable to different
computer architectures, and although well-written Fortran is more
efficient on vector machines, for almost all other situations they are
of similar speed, given equally good compilers.  The use of vector
machines (such as the Hitachi SR2201) was not anticipated when this
work on DPD began several years ago.  Furthermore, it was not believed
that the basic algorithm would vectorize well, due to the short vector
length in typical computations.  Large programs are easier to maintain
in C/C++ than in Fortran, although the increasingly well-supported
Fortran~90 and 95 make the difference less significant.

Finally, we comment on our findings in tuning the message passing
interface (MPI) calls for the Cray T3D\@.  In our simulations, it was
found that blocking calls (sends and receives) were faster than
non-blocking calls and were easier to use correctly.  Furthermore,
better scaling was achieved by sending the size of a variable-size
message in a separate message rather than probing incoming messages to
determine their size.  Finally, using derived data types to remove
unneeded data from messages was slower than sending everything.



\begin{table}[htbp]
  \begin{center}
    \begin{tabular}{cc}
      \tableline
      Model                  &       \\
      parameter              & Value \\
      \tableline
      $\alpha$               & 7.063 \\
      $\gamma$               & 5.650 \\
      $\sigma$               & 1.290 \\
      $m_i$                  & 1     \\
      $r_c$                  & 1.3   \\
      $\rho$                 & 4     \\
      \tableline
    \end{tabular}
  \end{center}
  \caption{Model parameters for the viscosity measurements of the
      lower-viscosity fluid.}
  \label{LowerViscosityParameters}
\end{table}

\begin{table}[htbp]
  \begin{center}
    \begin{tabular}{cc}
      \tableline
      Model                  &       \\
      parameter              & Value \\
      \tableline
      $\alpha$               & 7.063 \\
      $\gamma$               & 56.50 \\
      $\sigma$               & 4.079 \\
      $m_i$                  & 1     \\
      $r_c$                  & 1.3   \\
      $\rho$                 & 4     \\
      \tableline
    \end{tabular}
  \end{center}
  \caption{Model parameters for the viscosity measurements of the
      higher-viscosity fluid.}
  \label{HigherViscosityParameters}
\end{table}

\begin{table}[htbp]
  \begin{center}
    \begin{tabular}{cc}
      \tableline
      Model                  &       \\
      parameter              & Value \\
      \tableline
      $\alpha_0$             & 7.063 \\
      $\alpha_1$             & 7.487 \\
      $\gamma_0$             & 5.650 \\
      $\gamma_1$             & 56.50 \\
      $\sigma_0$             & 1.290 \\
      $\sigma_1$             & 4.079 \\
      $m_i$                  & 1     \\
      $r_c$                  & 1.3   \\
      $\rho$                 & 4     \\
      \tableline
    \end{tabular}
  \end{center}
  \caption{Model parameters used for the differing-viscosity
      immiscible fluid simulations.}
  \label{DifferingModelParameters}
\end{table}

\begin{table}[tbhp]
  \begin{tabular}{ccc}
    \tableline
    $\phi$ & Approximate range of $log_{10}t$ & $\beta$ \\
    \tableline
    0.2 & $0.17 \pm 0.09$ to $1.28 \pm 0.09$ & $0.092 \pm 0.014$ \\
    0.3 & $0.09 \pm 0.04$ to $1.07 \pm 0.03$ & $0.083 \pm 0.008$ \\
    0.4 & $0.13 \pm 0.05$ to $1.13 \pm 0.04$ & $0.068 \pm 0.006$ \\
    0.5 & $0.13 \pm 0.05$ to $1.05 \pm 0.06$ & $0.058 \pm 0.009$ \\
    0.6 & $0.15 \pm 0.05$ to $1.01 \pm 0.09$ & $0.042 \pm 0.010$ \\
    0.7 & $0.24 \pm 0.05$ to $1.25 \pm 0.07$ & $0.058 \pm 0.010$ \\
    0.8 & $0.16 \pm 0.13$ to $1.25 \pm 0.11$ & $0.031 \pm 0.014$ \\
    0.9 & $0.02 \pm 0.05$ to $1.11 \pm 0.12$ & $-0.24 \pm 0.07$ \\
    \tableline
    0.1 & $0.26 \pm 0.19$ to $2.57 \pm 0.11$ & $0.291 \pm 0.016$ \\
    0.2 & $0.53 \pm 0.13$ to $2.63 \pm 0.08$ & $0.335 \pm 0.008$ \\
    0.3 & $0.72 \pm 0.05$ to $2.51 \pm 0.07$ & $0.333 \pm 0.012$ \\
    0.4 & $0.72 \pm 0.06$ to $2.45 \pm 0.08$ & $0.341 \pm 0.016$ \\
    0.5 & $0.83 \pm 0.09$ to $2.47 \pm 0.06$ & $0.336 \pm 0.016$ \\
    0.6 & $0.92 \pm 0.11$ to $2.52 \pm 0.07$ & $0.336 \pm 0.012$ \\
    0.7 & $0.95 \pm 0.04$ to $2.65 \pm 0.06$ & $0.316 \pm 0.010$ \\
    0.8 & $1.03 \pm 0.10$ to $2.66 \pm 0.07$ & $0.299 \pm 0.013$ \\
    0.9 & $0.55 \pm 0.08$ to $2.56 \pm 0.13$ & $0.20 \pm 0.02$ \\
    \tableline
  \end{tabular}
  \bigskip
  \caption{Scaling exponents for two-dimensional differing-viscosity
    fluids as a function of viscous phase fraction, divided into
    ``early'' and ``late time''.}\label{glassy}
\end{table}

\begin{table}[tbhp]
  \begin{tabular}{ccc}
    \tableline
    $\phi$ & Approximate range of $log_{10}t$ & $\beta$ \\
    \tableline
    0.1 & $0.02 \pm 0.05$ to $1.74 \pm 0.05$ & $0.023 \pm 0.002$ \\
    0.2 & $-0.012 \pm 0.015$ to $1.12 \pm 0.04$ & $0.0219 \pm 0.0014$ \\
    0.3 & $0$             to $0.92 \pm 0.03$ & $0.0171 \pm 0.0017$ \\
    0.4 & $0.02 \pm 0.04$ to $0.88 \pm 0.05$ & $0.025 \pm 0.005$ \\
    0.5 & $0.00 \pm 0.02$ to $0.87 \pm 0.04$ & $0.028 \pm 0.007$ \\
    0.6 & $-0.008 \pm 0.016$ to $1.00 \pm 0.02$ & $0.0273 \pm 0.0012$ \\
    0.7 & $0.06 \pm 0.05$ to $1.03 \pm 0.04$ & $0.022 \pm 0.003$ \\
    0.8 & $-0.004 \pm 0.012$ to $1.31 \pm 0.04$ & $0.0228 \pm 0.0013$ \\
    0.9 & $0.06 \pm 0.08$ to $1.95 \pm 0.09$ & $0.021 \pm 0.003$ \\
    \tableline
    0.1 & $1.47 \pm 0.07$ to $2.62 \pm 0.07$ & $0.53  \pm 0.05$ \\
    0.2 & $0.78 \pm 0.04$ to $2.67 \pm 0.05$ & $0.377 \pm 0.008$ \\
    0.3 & $0.79 \pm 0.13$ to $2.67 \pm 0.08$ & $0.376 \pm 0.010$ \\
    0.4 & $0.71 \pm 0.16$ to $2.68 \pm 0.06$ & $0.360 \pm 0.007$ \\
    0.5 & $0.65 \pm 0.05$ to $2.68 \pm 0.05$ & $0.358 \pm 0.007$ \\
    0.6 & $0.703 \pm 0.019$ to $2.64 \pm 0.08$ & $0.364 \pm 0.007$ \\
    0.7 & $0.85 \pm 0.06$ to $2.62 \pm 0.10$ & $0.363 \pm 0.012$ \\
    0.8 & $1.08 \pm 0.04$ to $2.51 \pm 0.06$ & $0.406 \pm 0.017$ \\
    0.9 & $1.76 \pm 0.08$ to $2.70 \pm 0.03$ & $0.52 \pm 0.02$ \\
    \tableline
  \end{tabular}
  \bigskip
  \caption{Scaling exponents for three-dimensional differing-viscosity
    fluids as a function of viscous phase fraction, divided into
    ``early'' and ``late time''.}\label{3d-glassy}
\end{table}

\begin{table}[tbhp]
  \begin{tabular}{ccc}
    \tableline
    $\phi$ & Approximate range of $log_{10}t$ & $\beta$ \\
    \tableline
    0.2  & $0$             to $1.10 \pm 0.04$ & $0.123 \pm 0.016$ \\
    0.25 & $0.02 \pm 0.04$ to $1.04 \pm 0.06$ & $0.13 \pm 0.02$ \\
    0.3  & $0.02 \pm 0.08$ to $1.00 \pm 0.05$ & $0.135 \pm 0.012$ \\
    0.4  & $-0.015 \pm 0.010$ to $0.85 \pm 0.03$ & $0.114 \pm 0.017$ \\
    0.5  & $0.00 \pm 0.03$ to $0.90 \pm 0.08$ & $0.13 \pm 0.03$ \\
    \tableline
    0.05 & $0.19 \pm 0.19$ to $2.83 \pm 0.14$ & $0.283 \pm 0.019$ \\
    0.1  & $0.12 \pm 0.16$ to $2.81 \pm 0.18$ & $0.304 \pm 0.010$ \\
    0.15 & $0.24 \pm 0.20$ to $2.71 \pm 0.18$ & $0.304 \pm 0.010$ \\
    0.2  & $0.52 \pm 0.24$ to $2.71 \pm 0.15$ & $0.337 \pm 0.012$ \\
    0.25 & $1.0 \pm 0.3$   to $2.76 \pm 0.13$ & $0.39 \pm 0.04$ \\
    0.3  & $0.71 \pm 0.15$ to $2.53 \pm 0.22$ & $0.367 \pm 0.016$ \\
    0.4  & $1.11 \pm 0.16$ to $2.5 \pm 0.2$   & $0.415 \pm 0.019$ \\
    0.5  & $1.4 \pm 0.3$   to $2.50 \pm 0.09$ & $0.47 \pm 0.04$ \\
    \tableline
  \end{tabular}
  \bigskip
  \caption{Scaling exponents for two-dimensional equal-viscosity
    fluids as a function of minority phase fraction, divided into
    ``early'' and ``late time''.}\label{same-viscosity}
\end{table}

\begin{table}[tbhp]
  \begin{tabular}{ccc}
    \tableline
    $\phi$ & Approximate range of $log_{10}t$ & $\beta$ \\
    \tableline
    0.1 & $0$             to $1.41 \pm 0.07$ & $0.028 \pm 0.013$ \\
    0.2 & $0.01 \pm 0.04$ to $0.90 \pm 0.04$ & $0.026 \pm 0.007$ \\
    0.3 & $0.00 \pm 0.04$ to $0.73 \pm 0.05$ & $0.026 \pm 0.007$ \\
    0.4 & $0$             to $0.61 \pm 0.05$ & $0.023 \pm 0.005$ \\
    0.5 & $0$             to $0.60 \pm 0.05$ & $0.029 \pm 0.006$ \\
    \tableline
    0.1 & $1.35 \pm 0.11$ to $2.48 \pm 0.18$ & $0.43  \pm 0.08$ \\
    0.2 & $0.52 \pm 0.04$ to $2.68 \pm 0.02$ & $0.362 \pm 0.008$ \\
    0.3 & $0.43 \pm 0.04$ to $2.53 \pm 0.04$ & $0.366 \pm 0.009$ \\
    0.4 & $0.38 \pm 0.02$ to $2.47 \pm 0.03$ & $0.369 \pm 0.006$ \\
    0.5 & $0.32 \pm 0.04$ to $2.48 \pm 0.02$ & $0.364 \pm 0.007$ \\
    \tableline
  \end{tabular}
  \bigskip
  \caption{Scaling exponents for three-dimensional equal-viscosity
    fluids as a function of minority phase fraction, divided into
    ``early'' and ``late time''.}\label{3d-same}
\end{table}

\begin{table}[htbp]
  \begin{center}
    \begin{tabular}{lccc}
      \tableline
              & Number   & Elapsed time & Parallel \\
      Machine & of nodes & per node     & efficiency \\
      \tableline
      DEC 3000/400 AXP   & 1 & 624 & \\
      (133~MHz Alpha EV4) &&&\\
      \tableline
      Linux PC           & 1 & 180 & \\
      (350~MHz Intel Pentium II) &&&\\
      \tableline
      SGI Indigo2        & 1 & 200 & \\
      (195~MHz MIPS R10000) &&&\\
      \tableline
      EPCC Cray T3D        
      &   1 & 1254 & 1.00 \\
      (512 $\times{}$ 150~MHz Alpha EV4)
      &   2 &  575 & 1.09 \\
      &   4 &  354 & 0.89 \\
      &   8 &  251 & 0.62 \\
      &  16 &  206 & 0.38 \\
      &  32 &  214 & 0.18 \\
      \tableline
      HPCF Hitachi SR2201  
      &  1 & 1202 & 1.00 \\
      (256 $\times$ 150~MHz HARP-1E) 
      &  2 &  634 & 0.95 \\
      &  4 &  371 & 0.81 \\
      &  8 &  255 & 0.59 \\
      & 16 &  212 & 0.35 \\
      & 32 &  243 & 0.15 \\
      \tableline
    \end{tabular}
  \end{center}
  \caption{Elapsed time (in seconds) per node and parallel efficiency
      of various computers for the first 1000 time steps of a 6400
      particle simulation in two dimensions.  These computers were
      used to calculate the results in this paper.}\label{hpc-small1}
\end{table}

\begin{table}[htbp]
  \begin{center}
    \begin{tabular}{lccc}
      \tableline
              & Number   & Elapsed time & Parallel \\
      Machine & of nodes & per node     & efficiency \\
      \tableline
      SGI Octane        
      &   1 & 152 & 1.00 \\
      (250~MHz MIPS R10000)
      &   2 &  87 & 0.88 \\
      \tableline
      CSAR Cray T3E-1200E        
      &   1 & 143 & 1.00 \\
      (576 $\times{}$ 600~MHz Alpha EV5)
      &   2 &  96 & 0.74 \\
      &   4 &  75 & 0.48 \\
      &   8 &  67 & 0.27 \\
      &  16 &  70 & 0.13 \\
      \tableline
      CSAR SGI Origin2000
      &   1 & 133 & 1.00 \\
      (16 $\times{}$ 250~MHz MIPS R10000)
      &   2 &  81 & 0.83 \\
      &   4 &  69 & 0.49 \\
      &   8 &  60 & 0.28 \\
      \tableline
    \end{tabular}
  \end{center}
  \caption{Elapsed time (in seconds) per node and parallel efficiency
      of various computers for the first 1000 time steps of a 6400
      particle simulation in two dimensions.  These computers were not
      used to calculate the results in this paper.}\label{hpc-small2}
\end{table}


\begin{thebibliography}{10}

\bibitem{cov-nov:binary}
P.~V. Coveney and K.~E. Novik, Phys.\ Rev.\ E {\bf 54}, 5134 (1996).

\bibitem{cov-nov:erratum}
P.~V. Coveney and K.~E. Novik, Phys.\ Rev.\ E {\bf 55}, 4831 (1997).

\bibitem{nov-cov:binary}
K.~E. Novik and P.~V. Coveney, Int.\ J. Mod.\ Phys.\ C {\bf 8}, 909 (1997).

\bibitem{novik:finite-difference}
K.~E. Novik and P.~V. Coveney, J. Chem.\ Phys. {\bf 109}, 7667 (1998).

\bibitem{jury:scaling}
S.~I. Jury, P.~Bladon, S.~Krishna, and M.~E. Cates, Phys.\ Rev.\ E {\bf 59},
  R2535 (1999).

\bibitem{velasco:spinodal}
E.~Velasco and S.~Toxvaerd, Phys.\ Rev.\ Lett. {\bf 71}, 388 (1993).

\bibitem{velasco:off-critical}
E.~Velasco and S.~Toxvaerd, J. Phys.\ Condens.\ Matter {\bf 6}, A205 (1994).

\bibitem{velasco:separation}
E.~Velasco and S.~Toxvaerd, Phys.\ Rev.\ E {\bf 54}, 605 (1996).

\bibitem{laradji:spinodal}
M.~Laradji, S.~Toxvaerd, and O.~G. Mouritsen, Phys.\ Rev.\ Lett. {\bf 77}, 2253
  (1996).

\bibitem{kumar:novel}
P.~B.~S. Kumar and M.~Rao, Phys.\ Rev.\ Lett. {\bf 77}, 1067 (1996).

\bibitem{bib:so}
A.~Shinozaki and Y.~Oono, Phys.\ Rev.\ Lett. {\bf 66}, 173 (1991).

\bibitem{bib:so2}
A.~Shinozaki and Y.~Oono, Phys.\ Rev.\ E {\bf 48}, 2622 (1993).

\bibitem{bib:fv}
J.~E. Farrell and O.~T. Valls, Phys.\ Rev.\ B {\bf 40}, 7027 (1989).

\bibitem{bib:vf}
O.~T. Valls and J.~E. Farrell, Phys.\ Rev.\ E {\bf 47}, R36 (1993).

\bibitem{bib:walc}
Y.~Wu, F.~J. Alexander, T.~Lookman, and S.~Chen, Phys.\ Rev.\ Lett. {\bf 74},
  3852 (1995).

\bibitem{bib:lwac}
T.~Lookman, Y.~Wu, F.~J. Alexander, and S.~Chen, Phys.\ Rev.\ E {\bf 53}, 5513
  (1996).

\bibitem{corberi:early}
F.~Corberi, A.~Coniglio, and M.~Zannetti, Phys.\ Rev.\ E {\bf 51}, 5469 (1995).

\bibitem{bib:ctg}
A.~Chakrabarti, R.~Toral, and J.~D. Gunton, Phys.\ Rev.\ B {\bf 39}, 4386
  (1989).

\bibitem{chakrabarti:off-critical}
A.~Chakrabarti, R.~Toral, and J.~D. Gunton, Phys.\ Rev.\ E {\bf 47}, 3025
  (1993).

\bibitem{bib:rk}
D.~H. Rothman and J.~M. Keller, J. Stat.\ Phys. {\bf 52}, 1119 (1988).

\bibitem{rothman:spinodal}
D.~H. Rothman, Phys.\ Rev.\ Lett. {\bf 65}, 3305 (1990).

\bibitem{rothman:review}
D.~H. Rothman and S.~Zaleski, Rev.\ Mod.\ Phys. {\bf 66}, 1417 (1994).

\bibitem{bib:bal}
S.~Bastea and J.~L. Lebowitz, Phys.\ Rev.\ E {\bf 52}, 3821 (1995).

\bibitem{bib:appert}
C.~Appert, J.~F. Olson, D.~H. Rothman, and S.~Zaleski, J. Stat.\ Phys. {\bf
  81}, 181 (1995).

\bibitem{emerton:microemulsions}
A.~N. Emerton, P.~V. Coveney, and B.~M. Boghosian, Phys.\ Rev.\ E {\bf 55}, 708
  (1997).

\bibitem{emerton:erratum}
A.~N. Emerton, P.~V. Coveney, and B.~M. Boghosian, Phys.\ Rev.\ E {\bf 56},
  1286 (1997).

\bibitem{weig:minority}
F.~W.~J. Weig, P.~V. Coveney, and B.~M. Boghosian, Phys.\ Rev.\ E {\bf 56},
  6877 (1997).

\bibitem{bib:acg}
F.~J. Alexander, S.~Chen, and D.~W. Grunau, Phys.\ Rev.\ B {\bf 48}, 634
  (1993).

\bibitem{chen:late}
S.~Chen and T.~Lookman, J. Stat.\ Phys. {\bf 81}, 223 (1995).

\bibitem{osborn:lb}
W.~R. Osborn, E.~Orlandini, M.~R. Swift, J.~M. Yeomans, and J.~R. Banavar,
  Phys.\ Rev.\ Lett. {\bf 75}, 4031 (1995).

\bibitem{orlandini:structured}
E.~Orlandini, G.~Gonnella, and J.~M. Yeomans, Physica A {\bf 240}, 277 (1997).

\bibitem{gonnella:lamellar}
G.~Gonnella, E.~Orlandini, and J.~M. Yeomans, Phys.\ Rev.\ Lett. {\bf 78}, 1695
  (1997).

\bibitem{gonnella:droplet}
G.~Gonnella, E.~Orlandini, and J.~M. Yeomans, Phys.\ Rev.\ E {\bf 58}, 480
  (1998).

\bibitem{wagner:breakdown}
A.~J. Wagner and J.~M. Yeomans, Phys.\ Rev.\ Lett. {\bf 80}, 1429 (1998).

\bibitem{sappelt:glassy}
D.~Sappelt and J.~J{\"a}ckle, Europhys.\ Lett. {\bf 37}, 13 (1997).

\bibitem{hoogerbrugge:dpd}
P.~J. Hoogerbrugge and J.~M. V.~A. Koelman, Europhys.\ Lett. {\bf 19}, 155
  (1992).

\bibitem{espanol:dpd}
P.~Espa{\~n}ol and P.~Warren, Europhys.\ Lett. {\bf 30}, 191 (1995).

\bibitem{espanol:chain}
P.~Espa{\~n}ol, Phys.\ Rev.\ E {\bf 53}, 1572 (1996).

\bibitem{espanol:coarse}
P.~Espa{\~n}ol, M.~Serrano, and I.~Z{\'u}{\~n}iga, Int.\ J. Mod.\ Phys.\ C {\bf
  8}, 899 (1997).

\bibitem{groot:bridging}
R.~D. Groot and P.~B. Warren, J. Chem.\ Phys. {\bf 107}, 4423 (1997).

\bibitem{gardiner:stochastic}
C.~W. Gardiner, {\em Handbook of Stochastic Methods\/} (Springer, Berlin,
  1983).

\bibitem{risken:fokker}
H.~Risken, {\em The {F}okker--{P}lanck Equation: Methods of Solution and
  Applications\/}, 2nd edn. (Springer, Berlin, 1989).

\bibitem{espanol:hydro}
P.~Espa{\~n}ol, Phys.\ Rev.\ E {\bf 52}, 1734 (1995).

\bibitem{coveney:multicomponent}
P.~V. Coveney and P.~Espa{\~n}ol, J. Phys.\ A {\bf 30}, 779 (1997).

\bibitem{flekkoy:md}
E.~G. Flekk{\o}y and P.~V. Coveney, Phys.\ Rev.\ Lett. {\bf 83}, 1775 (1999).

\bibitem{flekkoy:dpd}
E.~G. Flekk{\o}y and P.~V. Coveney, Foundations of dissipative particle
  dynamics, {u}npublished.

\bibitem{bray:ordering}
A.~J. Bray, Adv.\ Phys. {\bf 43}, 357 (1994).

\bibitem{bastea:comment}
S.~Bastea and J.~L. Lebowitz, Phys.\ Rev.\ Lett. {\bf 75}, 3776 (1995).

\bibitem{siggia:late}
E.~D. Siggia, Phys.\ Rev.\ A {\bf 20}, 595 (1979).

\bibitem{fratzl:vacancy}
P.~Fratzl and O.~Penrose, Phys.\ Rev.\ B {\bf 50}, 3477 (1994).

\bibitem{tanaka:double}
H.~Tanaka and T.~Araki, Phys.\ Rev.\ Lett. {\bf 81}, 389 (1998).

\bibitem{kendon:spinodal}
V.~Kendon, J.-C. Desplat, P.~Bladon, and M.~E. Cates, Phys.\ Rev.\ Lett. {\bf
  83}, 576 (1999).

\bibitem{grant:spinodal}
M.~Grant and K.~R. Elder, Phys.\ Rev.\ Lett. {\bf 82}, 14 (1999).

\bibitem{landau:fluid}
L.~D. Landau and E.~M. Lifshitz, {\em Fluid Mechanics\/}, vol.~6 of {\em Course
  of Theoretical Physics\/} (Pergamon Press, Oxford, 1959).

\bibitem{allen:liquids}
M.~P. Allen and D.~J. Tildesley, {\em Computer Simulation of Liquids\/}
  (Clarendon, Oxford, 1987).

\bibitem{egelstaff:liquid}
P.~A. Egelstaff, {\em An Introduction to the Liquid State\/} (Clarendon,
  Oxford, 1992).

\bibitem{press:nr}
W.~H. Press, S.~A. Teukolsky, W.~T. Vetterling, and B.~P. Flannery, eds., {\em
  Numerical Recipes in C: The Art of Scientific Computing\/}, 2nd edn.
  (Cambridge University Press, Cambridge, 1992).

\bibitem{bromba:smoothing}
M.~U.~A. Bromba and H.~Ziegler, Anal.\ Chem. {\bf 53}, 1583 (1981).

\bibitem{marsh:properties}
C.~A. Marsh, G.~Backx, and M.~H. Ernst, Phys.\ Rev.\ E {\bf 56}, 1676 (1997).

\bibitem{marsh:thesis}
C.~Marsh, Theoretical aspects of dissipative particle dynamics, Ph.D. thesis,
  University of Oxford (1998).

\bibitem{swift:nonideal}
M.~R. Swift, W.~R. Osborn, and J.~M. Yeomans, Phys.\ Rev.\ Lett. {\bf 75}, 830
  (1995).

\end{thebibliography}
\end{document}